\documentclass[12pt]{article}
\usepackage[paper=a4paper,margin=1in]{geometry}
\usepackage[latin1]{inputenc}
\usepackage[T1]{fontenc} 
\usepackage{amsmath}
\usepackage{amsfonts}
\usepackage{amssymb}
\usepackage{amsthm}
\usepackage{graphicx}
\usepackage{mathrsfs}  

\newcommand{\ua}{\underline a \,}
\newcommand{\ub}{\underline b \,}

\numberwithin{equation}{section}

\begin{document}
\bibliographystyle{unsrt}

\title{Towards the quasi-localization of canonical GR\footnote{Talk 
given at the Conference on Recent Results in Mathematical Relativity, 
The Erwin Schr\"odinger Institute, Vienna, 2008 August 20-21, and 
dedicated to Bobby Beig on the occasion of his 60th birthday.}}

\author{ L\'aszl\'o B Szabados \\
Research Institute for Particle and Nuclear Physics \\
H-1525 Budapest 114, P. O. Box 49, Hungary }
\maketitle

\begin{abstract}
A general framework for a systematic quasi-localization of canonical 
general relativity and a new ingredient, the requirement of the 
gauge invariance of the boundary terms appearing in the calculation 
of Poisson brackets, are given. As a consequence of this it is shown, 
in particular, that the generator vector fields (built from the lapse 
and shift) of the quasi-local quantities must be divergence free with 
respect to a Sen-type connection; and the volume form induced from the 
spatial metric on the boundary surface must be fixed.

\end{abstract}


\section{Introduction}
\label{sec-1}

Conserved quantities have always had distinguished role in physics. 
Though for a general system no systematic way of finding them is 
known, but for systems whose dynamics can be described by a Hamiltonian 
in the canonical framework there is a way. The first systematic 
investigation of Einstein's general relativity (GR) in its canonical 
form was done in the ADM variables and was focused on asymptotically 
flat configurations \cite{ADM}. One of the key objects in the canonical 
formulation of the vacuum general relativity is the constraint function 
(`parameterized' by a function $N$ and a vector field $N^a$ on the 
manifold $\Sigma$, called the lapse and the shift, respectively): 

\begin{eqnarray}
C\bigl[N,N^e\bigr]:=-\int_\Sigma\Bigl\{\!\!\!\!&{}\!\!\!\!&\frac{1}{2
 \kappa}\Bigl(R-2\lambda+\frac{4\kappa^2}{\vert h\vert}\bigl[
 \frac{1}{n-1}\tilde p^2-\tilde p_{ab}\tilde p^{ab}\bigr]\Bigr)N
 \sqrt{\vert h\vert}+ \nonumber \\
\!\!\!\!&{}\!\!\!\!&+\bigl(2D_a\tilde p^{ab}\bigr)h_{bc}N^c\Bigr\}
 {\rm d}^nx. \label{eq:1.1}
\end{eqnarray}
Here the canonical variables are the fields $h_{ab}$ and $\tilde p
^{ab}$ on the connected $n$-manifold $\Sigma$, $h_{ab}$ being the 
(negative definite) spatial metric, $D_e$ is the corresponding 
Levi-Civita covariant derivative, $R$ is its curvature scalar and 
$\kappa:=8\pi G$ with Newton's gravitational constant $G$, and we 
allow a nontrivial cosmological constant $\lambda$ to be present. 
(Though primarily we are interested in the physical (3+1) dimensional 
case, the analysis can be done in $(n+1)$ dimensions without any extra 
effort, but $n\geq2$.) 

One of the basic observations of Arnowitt, Deser and Misner is that 
the constraint functions play a double role in the dynamics of GR (see 
also \cite{Ku}). In fact, the \emph{constraint} part of Einstein's 
equations is equivalent to $C[N,N^e]=0$ for every $N$ and $N^e$ and 
their \emph{formal} variational derivatives with respect to the 
canonical variables (given explicitly by 
(\ref{eq:2.2.a})-(\ref{eq:2.2.b})) appear in the canonical 
equations of motion, which are just the \emph{evolution parts} of the 
field equations: 

\begin{equation}
\dot h_{ab}=\frac{\delta H\bigl[N,N^e\bigr]}{\delta\tilde p^{ab}},
\hskip 20pt
\dot{\tilde p}{}^{ab}=-\frac{\delta H\bigl[N,N^e\bigr]}{\delta h
_{ab}}, \label{eq:1.2}
\end{equation}
where now $H[N,N^e]=C[N,N^e]$. Thus, apparently, it is the constraints 
that generate the evolution of the states of the theory through the 
canonical equations of motion in the phase space, i.e. the constraint 
functions appear to play the role of the Hamiltonian in the canonical 
formulation of general relativity. Another important observation of 
Arnowitt, Deser and Misner is that by the integral of the 
Landau--Lifshitz pseudotensor in an asymptotically Cartesian coordinate 
system, the total energy and linear momentum can be introduced, and 
these quantities turned out to be conserved during the time evolution 
of the system. 

However, as Regge and Teitelboim pointed out \cite{RT}, the constraint 
functions $C[N,N^e]$ with the $1/r$ and $1/r^2$ fall-off for the 
canonical variables are \emph{not} functionally differentiable in the 
strict sense (see e.g. \cite{wald}). The total variation of $C[N,N^e]$ 
with respect to $h_{ab}$ and $\tilde p^{ab}$ yields not only the 
expected volume terms, i.e. the \emph{formal} variational derivatives 
(\ref{eq:2.2.a})-(\ref{eq:2.2.b}) contracted with $\delta h_{ab}$ and 
$\delta\tilde p^{ab}$, respectively, but integrals on the boundary of 
$\Sigma$ at infinity as well (see equation (\ref{eq:2.1})). 
Thus, strictly speaking, the functional derivatives of $C[N,N^e]$ are 
the sums of smooth fields and \emph{distributions} concentrated on 
the boundary of $\Sigma$. Therefore, if we want to recover the correct 
evolution equation for the smooth tensor fields as the Hamiltonian 
equations of motion (rather than some distributional generalization of 
them), then the Hamiltonian must be functionally differentiable with 
respect to the canonical variables. Since adding a boundary integral 
to $C[N,N^e]$ does not change the formal functional derivatives, 
Regge and Teitelboim searched for the correct Hamiltonian in the form 

\begin{equation}
H\bigl[N,N^e\bigr]=C\bigl[N,N^e\bigr]+\frac{1}{\kappa}\oint_{\cal S}
B\bigl(N,N^e\bigr){\rm d}{\cal S}, \label{eq:1.3}
\end{equation}
where the integral on ${\cal S}$ is understood as the $r\rightarrow
\infty$ limit of the integrals on large spheres of radius $r$ in the 
asymptotically flat ends of $\Sigma$ and $B(N,N^e)$ is some expression 
of the canonical variables and a \emph{linear} expression of $N$ and 
$N^e$. They showed that there is, indeed, a boundary term which makes 
(\ref{eq:1.3}) differentiable. Moreover, as a bonus, for appropriately 
chosen $(N,N^e)$ this Hamiltonian automatically reproduces the total 
energy and linear momentum of Arnowitt, Deser and Misner as its value 
on the constraint surface and, in addition, the spatial angular 
momentum and centre-of-mass could also be introduced. Thus, the claim 
of a pure mathematical consistency yielded a physically highly desirable 
result. 

Nevertheless, as Beig and \'O Murchadha \cite{BM} showed, the asymptotic 
form of $N$ and $N^e$ should not be prescribed by hand. That is a 
consequence of the requirement of the compatibility of the boundary 
conditions and the evolution equations: Since the evolution equations 
involve the lapse and the shift, and they must preserve the boundary 
conditions imposed on the canonical variables, we get a restriction 
on the asymptotic form of $N$ and $N^e$. In fact, in the leading order 
they depend on the asymptotically Cartesian \emph{spatial} coordinates 
just like the Killing vectors of Minkowski spacetime. (On the other 
hand, the time dependence of $N$ and $N^e$ remained unrestricted.) 
Without this compatibility condition the initial data set, satisfying 
the boundary conditions at the initial instant, would evolve in the 
next instant into a new data set that would \emph{not} belong to the 
actual phase space, i.e. the evolution would take a data set out of 
the phase space. Beig and \'O Murchadha refined the Hamiltonian of 
Regge and Teitelboim, too, such that the new Hamiltonian is not only 
functionally differentiable, but finite on the whole phase space 
(rather than only on the constraint surface), and that these 
Hamiltonians form a Lie algebra with respect to the Poisson bracket 
as the Lie product. The constraints form a Lie ideal in this algebra, 
and their quotient, the algebra of observables, is isomorphic to the 
Poincar\'e algebra. The conserved quantities then become coordinates 
in this quotient algebra. 

Since the evolution equations specify only how the lapse and the 
shift depend asymptotically on the \emph{spatial} coordinates but 
not on the time coordinate, moreover the spatial angular momentum and 
the centre-of-mass of Beig and \'O Murchadha do not transform in the 
correct way under an asymptotic Poincar\'e transformation \emph{in 
the spacetime}, further refinements of the previous results were 
needed. As a resolution of these difficulties in \cite{szab03,szab06a} 
a distinction between the evolution vector fields that should be 
used in the Hamiltonian to generate the evolution of the states and 
the vector fields built from the lapse and the shift that should be 
used to define ADM type conserved quantities was made. The latter 
must be the asymptotic spacetime Killing fields, which turned out to 
be a special case of the former. 

In the present paper, these ideas will be applied to the case in which 
the manifold $\Sigma$ is compact with a smooth boundary 
${\cal S}:=\partial\Sigma$, i.e. at the quasi-local level (see e.g. 
\cite{szab04}). Thus, now we are interested in the Hamiltonian dynamics 
of general subsystems of the universe. The extension of the 
investigations of canonical general relativity to the quasi-local case 
is required by solving several problems. First, to have a deeper 
understanding of the (geometrical or thermodynamical) properties of 
black holes, for example to formulate the various (geometric) 
inequalities or the entropy bounds for black holes, the conserved 
quantities or, more generally, the observables of the gravitational 
`field' must be introduced quasi-locally. A further motivation of 
searching for quasi-local observables is the remarkable result of 
Torre \cite{To} that all the global observables for the vacuum 
gravitational field in a closed universe, built as spatial integrals 
of local functions of the initial data and their finitely many 
derivatives, are necessarily vanishing. Thus in closed universes we 
can associate non-trivial, locally constructible observables only 
with subsystems, bounded at one instant by some closed spacelike 
2-surface. Another motivation is the claim to see the content of the 
basic existence and uniqueness results of Friedrich and Nagy \cite{FN} 
for the initial-boundary value problem for the vacuum Einstein 
equations from the Hamiltonian point of view. 

Though the first few steps towards the systematic quasi-localization 
of canonical general relativity using the ADM variables have already 
been done \cite{szab06b}, the project remained incomplete and 
essential new ideas were needed. In the present paper, we continue 
these investigations. We refine our previous framework, and, in 
addition to the ideas introduced in the asymptotically flat context 
(and discussed above), only first principles (such as gauge 
invariance in every sense, covariance, etc), but no ad hoc ideas or 
elements (e.g. some a priori reference configuration, gauge choice or 
even implicitly given background structure) will be used.  We think 
that these would contradict the principle of equivalence (see also 
\cite{AG}) and hence the very spirit of Einstein's general relativity. 
It is the theory itself that should tell us what the boundary 
conditions, the observables, etc are, and we must read off these from 
the structure of the theory itself. For a different view (but an 
absolutely legitimate strategy), see e.g. \cite{CN}.

Though in the present paper we cannot complete the quasi-localization 
programme, we raise a new issue that should be discussed in connection 
with the quasi-localization of canonical GR. Namely, we argue that the 
boundary terms appearing in the calculation of the Poisson bracket of 
two Hamiltonians must be gauge invariant in every sense. We show that 
the requirement of this gauge invariance yields a restriction on the 
lapse and shift: the generator vector field built from them according 
to $K^a=Nt^a+N^a$, where $t^a$ is the future pointing unit timelike 
normal to the spacelike hypersurfaces $\Sigma$ in the spacetime, must 
be divergence free with respect to the Sen-type connection induced on 
the boundary, $\Delta_aK^a=0$. Remarkably enough, this is one of the 
ten (or, in $(n+1)$ dimensions, the $\frac{1}{2}(n+1)(n+2)$) spacetime 
Killing equations. A consequence of this condition is that the induced 
volume form $\varepsilon_{e_1\dots e_{n-1}}$ on the boundary ${\cal 
S}$ remains constant during the evolution, and hence it is natural to 
impose $\delta\varepsilon_{e_1\dots e_{n-1}}=0$. $\delta\varepsilon
_{e_1\dots e_{n-1}}=0$, as a part of the boundary conditions for the 
canonical variables, has already been found in special contexts, e.g. 
in connection with the functional differentiability of the constraint 
functions. The present investigations show that $\delta\varepsilon
_{e_1\dots e_{n-1}}=0$ should be a part of the `ultimate' boundary 
conditions, too. 

In the following section we formulate (and refine the previous 
attempts of) the quasi-localization programme of canonical GR and 
discuss the tools and the technical details that we need as well as 
the partial results. In section 3 we quasi-localize the canonical 
formulation of a single, real scalar field, and we will see that 
the boundary terms appearing in the Poisson bracket of two 
Hamiltonians should be interpreted as the energy-momentum and angular 
momentum flux, and hence a gauge invariant observable. 
In subsection \ref{sub-4.1}, we return to general relativity and 
study the analogous boundary terms in the Poisson bracket of two 
constraints. We derive here the boundary conditions $\Delta_aK^a=0$ 
and $\delta\varepsilon_{e_1\dots e_{n-1}}=0$. Finally, in subsection 
\ref{sub-4.2}, we find arguments both in favour of and against a 
Hamiltonian boundary term which is the basis of several suggestions 
for the quasi-local energy-momentum. 
Though there are a number of key issues even in connection with the 
present partial results (e.g. how they are related to the maximal 
dissipative boundary condition of Friedrich and Nagy \cite{FN}), 
we should leave these to a future investigation. 

Our notations and conventions are essentially those that were used in 
\cite{szab06b}. In particular, we use the abstract index formalism, 
the signature of the spacetime metric is $1-n$, and the curvature is 
defined by $-R^a{}_{bcd}X^b:=(D_cD_d-D_dD_c)X^a$. The analysis is 
based on certain formulae given explicitly in \cite{szab03,szab06b}. 
Our standard reference in canonical formalism is \cite{woodhouse}. 


\section{The quasi-localization programme}
\label{sec-2}

\subsection{The general programme}
\label{sub-2.1}

\subsubsection{The basic requirements}
\label{sub-2.1.1}

In the quasi-local canonical formulation of general relativity, the 
basic object is the quasi-local configuration space ${\cal Q}
(\Sigma)$ over the connected $n$-manifold $\Sigma$ with smooth 
boundary ${\cal S}:=\partial\Sigma$. This is the space of smooth 
(negative definite) metrics $h_{ab}$ on $\Sigma$ satisfying certain 
not-yet-specified boundary conditions on ${\cal S}$. The quasi-local 
phase space is defined to be its `cotangent bundle' $T^*{\cal Q}
(\Sigma)$, endowed with the natural symplectic structure. Thus the 
elements of the quasi-local phase space are the pairs $(h_{ab},
\tilde p^{ab})$ of fields, where $\tilde p^{ab}$ is a symmetric 
contravariant tensor density of weight one. 
($\tilde p^{ab}$ is usually interpreted as a 1-form at the point 
$h_{ab}\in{\cal Q}(\Sigma)$: if $h_{ab}(u)$ is any smooth 1-parameter 
family of metrics such that $h_{ab}(0)=h_{ab}$, i.e. $h_{ab}(u)$ is a 
`curve' in ${\cal Q}(\Sigma)$ through $h_{ab}$, then the tensor 
field $\delta h_{ab}$ on $\Sigma$, defined pointwise as $\delta h
_{ab}(p):=({\rm d}h_{ab}(p,u)/{\rm d}u)\vert_{u=0}$, $\forall p\in
\Sigma$, is called the tangent of the `curve' at $h_{ab}$. In fact, 
the directional derivative of any functionally differentiable function 
$F:{\cal Q}(\Sigma)\rightarrow\mathbb{R}$ along the `curve' at the 
point $h_{ab}$ is $\delta F:=({\rm d}F(h_{ab}(u))/{\rm d}u)\vert_{u=0}
=\int_\Sigma\frac{\delta F}{\delta h_{ab}}\delta h_{ab}{\rm d}^nx$, 
which defines the action of $\delta h_{ab}$ on any such $F$ and the 
natural pairing of $\delta h_{ab}$ and the `exact 1-form' $dF=
\frac{\delta F}{\delta h_{ab}}$ at the same time. Thus, it is natural 
to define the action of $\tilde p^{ab}$ on the `tangent vector' 
$\delta h_{ab}$ at the point $h_{ab}\in{\cal Q}(\Sigma)$ by the 
integral $\int_\Sigma\delta h_{ab}\tilde p^{ab}{\rm d}^nx$. Note that 
while in the asymptotically flat context the requirement of the 
finiteness of the integral $\int_\Sigma\delta h_{ab}\tilde p^{ab}
{\rm d}^nx$ together with the evolution equations restrict the 
fall-off rate $k$ of the metric, namely \cite{szab03} it must be 
$k\geq\frac{1}{2}(n-1)$, in the quasi-local case we obtain no 
restriction for the canonical variables.) 
Clearly, the differentiability of functions on the quasi-local phase 
space depends not only on the function itself, but on the boundary 
conditions that are imposed on the canonical variables in the 
definition of $T^*{\cal Q}(\Sigma)$. We stress that the boundary 
conditions are parts of the definition of the phase space. (For a more 
detailed discussion of this issue, see e.g. \cite{Ki}.) The canonical 
symplectic structure can be characterized equivalently by specifying 
the Poisson bracket of any two \emph{functionally differentiable} 
functions. For any such $G$ and $H:T^*{\cal Q}(\Sigma)\rightarrow
\mathbb{R}$, it is $\{G,H\}:=\int_\Sigma(\frac{\delta G}{\delta\tilde p
^{ab}}\frac{\delta H}{\delta h_{ab}}-\frac{\delta H}{\delta\tilde p
^{ab}}\frac{\delta G}{\delta h_{ab}}){\rm d}^nx$. 

In a more pedagogical approach the lapse and the shift are also 
considered to be configuration variables (see e.g. \cite{Beig}), and 
there are additional constraints (the `primary constraints') that 
the momenta conjugate to the lapse and the shift are vanishing. Then 
a systematic constraint analysis shows that this `big' phase space 
$T^*\tilde{\cal Q}(\Sigma)$ can always be reduced to $T^*{\cal Q}
(\Sigma)$ in a straightforward way, and the role of the lapse and the 
shift is reduced from dynamical variables only to `parameters'. In 
this sense, $T^*{\cal Q}(\Sigma)$ can also be considered as a 
`partially reduced' phase space. We save this reduction by starting 
with $T^*{\cal Q}(\Sigma)$ as the phase space.

Our ultimate aim is the quasi-localization of canonical general 
relativity, i.e. finding 

\begin{description}
\item{1.} a boundary term $B(N,N^e)$ (we will call its integral the 
          Hamiltonian boundary term), built from the canonical 
          variables and depending linearly on the lapse and the shift;
\item{2.} boundary conditions for the canonical variables $(h_{ab},
          \tilde p^{ab})$ on ${\cal S}$;
\item{3.} boundary conditions for the lapse $N$ and the shift $N^e$ on 
          ${\cal S}$ 
\end{description}
such that 

\begin{description}
\item{i.} the Hamiltonian $H:T^*{\cal Q}(\Sigma)\rightarrow\mathbb{R}$, 
          given by (\ref{eq:1.3}) and `parameterized' by lapses and 
          shifts satisfying the boundary conditions in point 3, is 
          functionally differentiable with respect to the canonical 
          variables;
\item{ii.} the evolution equations (\ref{eq:1.2}) with lapses and shifts 
          satisfying the boundary condition in point 3. above preserve 
          the boundary conditions imposed on the canonical variables in 
          point 2;
\item{iii.} the constraints close to a Poisson algebra ${\cal C}$ (which 
          we call the quasi-local constraint algebra);
\item{iv.} the value of the Hamiltonian on the constraint surface (i.e. 
          if $C[N,N^e]=0$) must be a $2+(n-1)$-covariant, gauge-invariant 
          expression of the boundary data on ${\cal S}$. 
\end{description}
Before turning to the mathematical realization of these requirements we 
should discuss these points.

\subsubsection{The discussion of the requirements}
\label{sub-2.1.2}

Clearly, requirement (i) is just that of Regge and Teitelboim \cite{RT}, 
and (ii) is taken from Beig and \'O Murchadha \cite{BM}. While the first 
is a generally accepted condition in the relativity community, 
the second is apparently not appreciated enough. Moreover, the usual 
(and in the quasi-local canonical approaches almost exclusively 
adopted) view is that the lapse and the shift on the boundary ${\cal 
S}$ should be arbitrary, just because they are thought of as the $n+1$ 
pieces of the spacetime evolution vector field, and their 
arbitrariness are thought to express our freedom to choose the fleet 
of observers in the spacetime as we wish. 

However, as we learnt from the structure of canonical GR of 
asymptotically flat spacetimes, requirement (ii) may yield non-trivial 
restrictions on the lapse and the shift. Moreover, as the study of 
the quasi-local conserved quantities of matter fields in Minkowski 
spacetime shows (and as we summarize the basic things in subsection 
\ref{sub-3.1}), we must make a distinction between the 
\emph{symmetries} (which define the conserved quantities) and the 
general \emph{evolution vector fields} (which are used to define the 
time evolution of the initial data set in the spacetime). The former 
is the solutions of a linear partial differential equation (and the 
number of their independent solutions is finite), but the latter 
is quite arbitrary. Thus, in general relativity, the lapse and shift 
parts of a general evolution vector field may be, but the lapse and 
shift parts of a vector field defining `conserved' quantities are 
probably restricted. In particular, as we will see in subsection 
\ref{sub-2.3.3}, the `most natural' quasi-local Hamiltonian cannot 
provide an acceptable expression for the energy-momentum if the lapse 
and shift on ${\cal S}$ are chosen independently of the canonical 
variables. We expect that these conditions are given only implicitly, 
possibly in the form of a system of \emph{linear} partial differential 
equations (p.d.e.) on ${\cal S}$. In fact, in section \ref{sec-4} we 
derive an equation for the lapse and shift which should be a part of 
such a system of linear p.d.e. 

Requirement (iii) expresses the claim that the gauge content of the 
theory must be represented in a correct way by the constraints even 
quasi-locally. The motivation of this expectation comes from the 
applicability of the standard canonical formalism (see e.g. 
\cite{woodhouse}). In fact, for $n\geq3$ general relativity is 
expected to have internal degrees of freedom, which are expected to 
be represented in the canonical framework by the points of the 
reduced phase space. However, strictly speaking the reduced phase 
space can be formed as the space of gauge orbits in the constraint 
surface only when the Hamiltonian vector fields of the constraint 
functions form an involutive distribution or, equivalently, if the 
constraint functions close to a Poisson algebra. 

Though in the asymptotically flat case the Hamiltonians of Beig and 
\'O Murchadha also close to a Poisson algebra (in which the 
constraints form an ideal) \cite{BM}, we will see in subsection 
\ref{sub-3.2.2} that quasi-locally this is not true even for the 
scalar field, and hence it cannot be expected in general relativity 
either. In fact, the boundary terms appearing in the Poisson bracket 
of two Hamiltonians (and which destroy the Poisson algebra structure) 
describe incoming and outgoing energy-momentum and angular momentum 
fluxes. Thus quasi-locally the algebra of observables is much bigger 
than the set of the Hamiltonians. 

In the present investigations, we do not use any `natural' (or rather 
ad hoc) conditions e.g. for the canonical variables. We use only 
first principles, such as gauge invariance (in any sense) or 
covariance. Requirement (iv) is the manifestation of this idea. In 
particular, the boundary term $B(N,N^e)$ should not depend on the 
normal directional derivatives of the lapse and the shift, or on 
the actual lapse and shift separately, but only on their 
spacetime-covariant combination $K^e:=Nt^e+N^e$. Moreover, it should 
not depend on the actual choice for the normals $(t^a,v^a)$ of 
${\cal S}$ but only on the intrinsic and extrinsic \emph{geometry} of 
${\cal S}$ (but not e.g. on the extension of the geometrical quantities 
off ${\cal S}$).

However, because of the strong interplay between the Hamiltonian 
boundary term and the boundary conditions both for the canonical 
variables and the lapse and shift, it would be extremely difficult to 
determine these \emph{three} unknown things in a single step. Indeed, 
in the traditional approach usually one of these three is a priori 
given, and by the requirements (i) and (ii) we can determine the 
other two in a relatively straightforward way. For example, in the 
asymptotically flat case we have a priori boundary conditions, from 
which Beig and \'O Murchadha could determine both the Hamiltonian 
boundary term and the boundary conditions for the lapse and shift. 
In subsection \ref{sub-2.3} we summarize the special cases in which 
various Hamiltonian boundary terms were a priori given, from which 
the boundary conditions could be determined. (The necessary tools and 
technical ingredients will be summarized in subsection \ref{sub-2.2}.)
Thus, in order to be able to manage the general problem, a new idea 
would be needed. 

As we already mentioned in the introduction, such a new idea could be 
the observation that the boundary terms in the Poisson bracket of two 
Hamiltonians (which we call the Poisson boundary term) should be the 
sum of the Hamiltonian boundary term and terms analogous to the 
energy-momentum and angular momentum flows between the system and the 
rest of the universe, and hence these must be gauge invariant. 
This idea is based on the moral of the analogous investigations of 
a scalar field, the subject of section 3.


\subsection{The main ingredients}
\label{sub-2.2}

\subsubsection{The variational formula}
\label{sub-2.2.1}

Though formally the manifold $\Sigma$ on which the canonical 
variables $(h_{ab},\tilde p^{ab})$ as fields are defined is an 
abstract $n$-manifold, it is convenient (and illuminating, and hence 
useful) to consider this as the typical leaf of a foliation $\{\Sigma
_t\}$ of (a piece of) the spacetime by spacelike hypersurfaces and 
identify $\Sigma$ at the coordinate time instant $t$ with 
$\Sigma_t$. Thus although in the phase-space context there is no 
reason to speak about the normal of $\Sigma$, in the spacetime we 
have a uniquely defined future pointing unit timelike normal $t^a$ 
of the leaves $\Sigma_t$ (and hence a projection $P^a_b:=\delta^a_b
-t^at_b$ of the spacetime tangent spaces to the tangent spaces of 
the leaves, too), and we frequently rewrite formulae given in the 
phase-space context to its spacetime form or vice versa freely if 
needed. For example, the spacetime form of the lapse and the shift 
is the so-called evolution vector field $\xi^a=Nt^a+N^a$, by means of 
which the spacetime form of the constraint function (\ref{eq:1.1}) 
is just the integral $C[\xi^e]:=\frac{1}{\kappa}\int_\Sigma\xi^a
({}^MG_{ab}+\lambda g_{ab})t^b{\rm d}\Sigma$. Here $g_{ab}$ is the 
spacetime metric, ${}^MG_{ab}$ is the corresponding Einstein tensor 
and the induced volume element on $\Sigma$ is ${\rm d}\Sigma:=
\frac{1}{n!}t^e\varepsilon_{ea_1...a_n}$. ($\varepsilon_{a_1...a
_{n+1}}$ is the volume $(n+1)$-form in the spacetime.) Note that the 
timelike unit normal to $\Sigma$ is globally well defined if the 
spacetime is time orientable. 

Also, we will need the expression of the canonical momentum in 
terms of the Lagrange variables, i.e. the configuration and velocity 
variables. The latter is defined with respect to some spacetime 
vector field $\xi^e$ such that the hypersurfaces $\Sigma_t$ of the 
foliation are obtained from $\Sigma$ by Lie dragging along the integral 
curves of $\xi^e$. We assume that these hypersurfaces are spacelike. 
Let $t^a$ be their future pointing unit timelike normal, the lapse and 
the shift parts of $\xi^e$ are $N:=t_e\xi^e$ and $N^e:=P^e_f\xi^f$, 
respectively, and the acceleration of the hypersurfaces is $a_e:=t^a
\nabla_at_e=-D_e\ln N$. Here $D_e$ denotes the induced Levi-Civita 
derivative operator on $T\Sigma_t$. The time derivative of a purely 
\emph{spatial} tensor field $T^{a...}_{b...}$ is defined to be the 
projection to the hypersurfaces of its Lie derivative along $\xi^e$, 
i.e. $\dot T^{a...}_{b...}:=(L_{\xi}T^{c...}_{d...})P^a_c...P^d_b...
=N(L_{\bf t}T^{c...}_{d...})P^a_c...P^d_b..+L_{\bf N}T^{a...}_{b...}$. 
In particular, $\dot h_{ab}=2N\chi_{ab}+L_{\bf N}h_{ab}$. Thus 
essentially the extrinsic curvature $\chi_{ab}$ of the leaves of the 
foliation plays the role of the velocity, by means of which the 
canonical momentum is known to be $\tilde p^{ab}=\frac{1}{2\kappa}
\sqrt{\vert h\vert}(\chi^{ab}-\chi h^{ab})$. Here $\chi$ is the $h
_{ab}$-trace of $\chi_{ab}$. 

To calculate the total variation of (\ref{eq:1.1}), let $N(u)$, 
$N^a(u)$, $h_{ab}(u)$ and $\tilde p^{ab}(u)$, $u\in(-\epsilon,
\epsilon)$, be any smooth 1-parameter families of lapses, shifts, 
metrics and canonical momenta, respectively. Then we define the 
corresponding variation of any of their function, $F=F(N,N^a,h_{ab},
\tilde p^{ab})$, to be its $u$-derivative at $u=0$, i.e. $\delta F
:=({\rm d}F(N(u),N^a(u),h_{ab}(u),$ $\tilde p^{ab}(u))/{\rm d}u)\vert
_{u=0}$. The corresponding variation of the constraint function 
$C[N,N^e]$, taken from \cite{szab03}, is 

\begin{eqnarray}
\delta C\bigl[N,N^e\bigr]\!\!\!\!&=\!\!\!\!&C\bigl[\delta N,\delta N^e
 \bigr]+\int_\Sigma\Bigl(\frac{\delta C[N,N^e]}{\delta h_{ab}}\delta 
 h_{ab}+\frac{\delta C[N,N^e]}{\delta\tilde p^{ab}}\delta\tilde p^{ab}
 \Bigr){\rm d}^nx+ \nonumber \\
+\frac{1}{2\kappa}\oint_{\partial\Sigma}\!\!\!\!&\Bigl\{\!\!\!\!&N
 \bigl(h^{ab}v^e(D_e\delta h_{ab})-v^a(D^b\delta h_{ab})\bigr)+\bigl(
 v^aD^bN-h^{ab}v^eD_eN\bigr)\delta h_{ab} \nonumber \\
\!\!\!\!&+\!\!\!\!&\frac{2\kappa}{\sqrt{\vert h\vert}}\bigl(2N^av_e
 \tilde p^{eb}-N^ev_e\tilde p^{ab}\bigr)\delta h_{ab}+4\kappa N_av_b
 \frac{\delta\tilde p^{ab}}{\sqrt{\vert h\vert}}\Bigr\}{\rm d}{\cal S}.
 \label{eq:2.1}
\end{eqnarray}
Here $v^a$ is the {\it outward} pointing unit normal of ${\cal S}$ in 
$\Sigma$, ${\rm d}{\cal S}$ is the induced volume element on the 
boundary, and the \emph{formal} variational derivatives are 

\begin{eqnarray}
\frac{\delta C[N,N^e]}{\delta h_{ab}}:\!\!\!\!&=\!\!\!\!&\frac{1}{2
 \kappa}\sqrt{\vert h\vert}\Bigl\{N\Bigl(R^{ab}-Rh^{ab}+2\lambda h
 ^{ab}+\frac{8\kappa^2}{\vert h\vert}\bigl(\tilde p^a{}_c\tilde p
 ^{cb}-\nonumber \\
\!\!\!\!&-\!\!\!\!&\frac{1}{n-1}h_{cd}\tilde p^{cd}\tilde p^{ab}
 \bigr)\Bigr)+D^aD^bN-h^{ab}D_cD^cN\Bigr\}-L_{\bf N}\tilde p^{ab}+
 \nonumber \\
\!\!\!\!&+\!\!\!\!&\frac{1}{4\kappa}Nh^{ab}\sqrt{\vert h\vert}\Bigl(
 R-2\lambda+\frac{4\kappa^2}{\vert h\vert}\bigl(\frac{1}{n-1}\tilde 
 p^2-\tilde p^{cd}\tilde p_{cd}\bigr)\Bigr), \label{eq:2.2.a} \\
\frac{\delta C[N,N^e]}{\delta\tilde p^{ab}}:\!\!\!\!&=\!\!\!\!&
 \frac{4\kappa}{\sqrt{\vert h\vert}}N\Bigl(\tilde p_{ab}-\frac{1}{n-1}
 \tilde p^{cd}h_{cd}h_{ab}\Bigr)+L_{\bf N}h_{ab}. \label{eq:2.2.b}
\end{eqnarray}
Here $R_{ab}$ is the Ricci tensor of $D_e$, and note that the last 
line of (\ref{eq:2.2.a}) is $-\frac{1}{2}h^{ab}$ times the integrand 
of the constraint function $C[N,0]$. 

Therefore, $C[N,N^e]$ is functionally differentiable (in the strict 
sense of \cite{wald}) with respect to the canonical variables only 
if the boundary integral in (\ref{eq:2.1}) is vanishing, in which 
case the functional derivatives themselves are given by 
(\ref{eq:2.2.a}) and (\ref{eq:2.2.b}). Then, provided the constraints 
are satisfied, the vacuum evolution equations (with the cosmological 
constant) are precisely the canonical equations of motion 
(\ref{eq:1.2}).

\subsubsection{A quick review of the geometry of the boundary surface}
\label{sub-2.2.2}

To evaluate the boundary terms in (\ref{eq:2.1}), it seems useful to 
split the variation of the metric $h_{ab}$ at the points of ${\cal 
S}$ with respect to the boundary. Moreover, in the subsequent 
subsections several expressions on ${\cal S}$, obtained originally 
in the traditional $n+1$ form, will have to be rewritten in a $2+
(n-1)$ form. Nevertheless, most of these notions have a non-trivial 
meaning only if $n\geq3$; and hence when we use them we assume that 
$n\geq3$. Thus now, in a nutshell, we summarize the basic geometric 
objects that we need in what follows. A more detailed discussion of 
these concepts is given e.g. in \cite{szab94,szab04}. 

To avid confusion, the Kronecker delta on $\Sigma$ will be denoted by 
$P^a_b$, and the $h_{ab}$-orthogonal projection to ${\cal S}$ is $\Pi
^a_b:=P^a_b+v^av_b$. Then the induced metric and the corresponding 
intrinsic Levi-Civita covariant derivative on ${\cal S}$ is $q_{ab}
:=h_{cd}\Pi^c_a\Pi^d_b$ and $\bar\delta_e$, respectively, and let us 
introduce another derivative operator simply by $\bar\Delta_e:=\Pi^f
_eD_f$. The extrinsic curvature of ${\cal S}$ in $\Sigma$ will be 
defined by $\nu_{ab}:=\Pi^c_a\Pi^d_bD_cv_d$. The difference of these 
two derivative operators is just the extrinsic curvature: $\bar
\Delta_eX^a=\bar\delta_e(\Pi^a_bX^b)-v^a\bar\delta_e(v_bX^b)+(\nu_{eb}
v^a-\nu_e{}^av_b)X^b$ for any $X^a$ tangent to $\Sigma$. The induced 
volume $(n-1)$ form and volume element on ${\cal S}$ are $\varepsilon
_{e_1...e_{n-1}}:=t^av^b\varepsilon_{abe_1...e_{n-1}}$ and ${\rm d}
{\cal S}:=\frac{1}{(n-1)!}t^av^b\varepsilon_{abe_1...e_{n-1}}$, 
respectively. Note that with these conventions, the Gauss theorem takes 
the form $\int_\Sigma D_aX^a{\rm d}\Sigma=-\oint_{\cal S}v_aX^a{\rm d}
{\cal S}$ for any vector field $X^a$ on $\Sigma$.

The boundary ${\cal S}=\partial\Sigma$ can be considered as a 
submanifold in the spacetime, too. In the spacetime context the 
induced metric is $q_{ab}=g_{cd}\Pi^c_a\Pi^d_b$, and the area 2-form 
on the 2-planes normal to ${\cal S}$ is ${}^\bot\varepsilon_{ab}:=
t_av_b-t_bv_a$. Here, both the projection $\Pi^a_b$ and the area 
2-form ${}^\bot\varepsilon_{ab}$ are independent of the actual choice 
of the normals $(t^a,v^a)$. Note that the normals are not specified 
by ${\cal S}$, but if ${\cal S}$ is considered to be the boundary of 
$\Sigma$, then they are chosen as in the previous paragraph. If 
${\cal S}$ is orientable and at least a neighbourhood of ${\cal S}$ 
in $M$ is space and time orientable, then $(t^a,v^a)$ can be chosen 
to be globally defined, yielding a global trivialization of the normal 
bundle $N{\cal S}$ of ${\cal S}$ in $M$. 
The two derivative operators $\Delta_e$ and $\delta_e$, acting on 
any Lorentzian $n+1$ vector field $X^a$, are defined by $\Delta_eX^a
:=\Pi^f_e\nabla_fX^a$ and $\delta_eX^a:=\Pi^a_b\Delta_e(\Pi^b_cX^c)
+(\delta^a_b-\Pi^a_b)\Delta_e((\delta^b_c-\Pi^b_c)X^c)$. The 
extrinsic curvature tensor of ${\cal S}$ in $M$ is $Q^a{}_{eb}:=-\Pi
^a_c\Delta_e\Pi^c_b=\tau^a{}_et_b-\nu^a{}_ev_b$, where $\tau_{ab}$ 
and $\nu_{ab}$ are the individual (symmetric) extrinsic curvatures of 
${\cal S}$ in $M$ corresponding to the unit normals $t_a$ and $v_a$, 
respectively. The corresponding traces are $\tau:=\tau_{ab}q^{ab}$ 
and $\nu:=\nu_{ab}q^{ab}$, respectively. 
The difference of the two derivative operators is this extrinsic 
curvature tensor: $\Delta_eX^a=\delta_eX^a+(Q^a{}_{eb}-Q_{be}{}^a)
X^b$. $\delta_e$ is not only the Levi-Civita derivative operator 
$\bar\delta_e$ on the tangent bundle of ${\cal S}$, but it acts on 
the normal bundle $N{\cal S}$ of ${\cal S}$, spanned by the two 
normals $t^a$ and $v^a$, as well. Its action can be characterized by 
the connection 1-form $A_e:=(\Delta_et_a)v^a=(\delta_et_a)v^a$. On 
the other hand, for vectors tangent to $\Sigma$ in $M$ the two 
derivative operators $\bar\Delta_e$ and $\Delta_e$ coincide, which 
fact will be used several times when we rewrite expressions given in 
the $n+1$ form into its $2+(n-1)$ form. 

At the points of ${\cal S}$, the splitting $h_{ab}=q_{ab}-v_av_b$ 
implies the variation $\delta h_{ab}=\delta q_{ab}-v_a\delta v_b-v_b
\delta v_a$. It is straightforward to determine the various 
projections of $\delta h_{ab}$ (for details see \cite{szab06b}). 
These are 

\begin{eqnarray}
&{}&\delta h_{cd}\Pi^c_a\Pi^d_b=\delta q_{cd}\Pi^c_a
 \Pi^d_b, \hskip 15pt
\delta h_{cd}v^c\Pi^d_b=-\delta v^aq_{ab}, 
 \nonumber  \\
&{}&\delta h_{cd}v^cv^d=2v^a\delta v_a=-2v_a\delta v^a. \label{eq:2.3}
\end{eqnarray}
Thus the independent variations can be represented by $\delta q_{cd}
\Pi^c_a\Pi^d_b$ and $\delta v^a$. 

The curvature of the connections $\delta_e$ and $\Delta_e$, 
respectively, are given by 

\begin{eqnarray}
f^a{}_{bcd}\!\!\!\!&=\!\!\!\!& -{}^\bot\varepsilon^a{}_b\bigl(
 \delta_cA_d-\delta_dA_c\bigr)+{}^{\cal S} R^a{}_{bcd}, 
 \label{eq:2.4.a} \\
F^a{}_{bcd}\!\!\!\!&=\!\!\!\!&f^a{}_{bcd}-\delta_c\bigl(Q^a{}_{db}-
 Q_{bd}{}^a\bigr)+\delta_d\bigl(Q^a{}_{cb}-Q_{bc}{}^a\bigr)+
 \nonumber\\
& &+Q^a{}_{ce}Q_{bd}{}^e+Q_{ec}{}^aQ^e{}_{db}-Q^a{}_{de}Q_{bc}{}^e
 -Q_{ed}{}^aQ^e{}_{cb}, \label{eq:2.4.b}
\end{eqnarray}%
where ${}^{\cal S}R^a{}_{bcd}$ is the curvature tensor of the 
intrinsic Levi-Civita connection $\bar\delta_e$ of $({\cal S},q
_{ab})$. Of course, for $n=3$, it can also be written in the form 
$\frac{1}{2}{}^{\cal S} R(\Pi^a_cq_{bd}-\Pi^a_dq_{bc})$, where ${}
^{\cal S} R$ is the curvature scalar. The curvature $F^a{}_{bcd}$ 
turns out to be just the pull-back to ${\cal S}$ of the spacetime 
curvature 2-form: $F^a{}_{bcd}={}^MR^a{}_{bef}\Pi^e_c\Pi^f_d$. Its 
various projections, 

\begin{eqnarray}
\Pi^e_a\Pi^f_bF_{efcd}\!\!\!\!&=\!\!\!\!&{}^{\cal S}R_{abcd}+
 \tau_{ac}\tau_{bd}-\tau_{ad}\tau_{bc}-\nu_{ac}\nu_{bd}+\nu_{ad}
 \nu_{bc}, \label{eq:2.5.a} \\
t^a\Pi^f_bF_{afcd}\!\!\!\!&=\!\!\!\!&\delta_c\tau_{db}-\delta_d
 \tau_{cb}+A_c\nu_{db}-A_d\nu_{cb}, \label{eq:2.5.b} \\
v^a\Pi^f_bF_{afcd}\!\!\!\!&=\!\!\!\!&\delta_c\nu_{db}-\delta_d
 \nu_{cb}+A_c\tau_{db}-A_d\tau_{cb}, \label{eq:2.5.c} \\
t^av^bF_{abcd}\!\!\!\!&=\!\!\!\!&\tau_{ec}\nu^e{}_d-\tau_{ed}\nu
 ^e{}_c+\delta_cA_d-\delta_dA_c, \label{eq:2.5.d} 
\end{eqnarray}
are the so-called Gauss, Codazzi--Mainardi, and Ricci equations.


\subsection{Special cases}
\label{sub-2.3}

\subsubsection{The quasi-local constraint algebra}
\label{sub-2.3.1}

A special, genuinely quasi-local case in which the programme could be 
completed is when there is no Hamiltonian boundary term, i.e. when 
we are interested in the boundary conditions both for the canonical 
variables and for the lapse and shift that make the \emph{constraint} 
functions functionally differentiable and close to a Poisson algebra. 
The significance of this special case is that the constraints 
represent those parts of the field equations that are expected to 
generate the gauge motions in phase space. Thus to understand the 
gauge content of GR at the quasi-local level, we should first clarify 
this special case. 

Decomposing the boundary terms in (\ref{eq:2.1}) with respect to the 
boundary ${\cal S}$ according to subsection \ref{sub-2.2.2}, we can 
read off the condition of the functional differentiability of $C[N,
N^e]$ \cite{szab06b}: the lapse and the shift are vanishing on 
${\cal S}$ and the induced volume $(n-1)$-form $\varepsilon_{e_1
\dots e_{n-1}}$ is fixed on ${\cal S}$. It is straightforward to 
check that the boundary condition $\delta\varepsilon_{e_1\dots e
_{n-1}}=0$ is preserved by the evolution equations with shifts and 
lapses vanishing on ${\cal S}$. Thus the quasi-local Hamiltonian 
phase space $T^*{\cal Q}(\Sigma)$ is split into disjoint sectors $T
^*{\cal Q}(\Sigma,\varepsilon_{e_1,\dots,e_{n-1}})$, labeled by the 
value of the volume form on ${\cal S}$. The constraint functions are 
differentiable in the directions tangential to these sectors, but not 
in the directions transversal to them. The Poisson bracket of any two 
constraint functions $C[N,N^e]$ and $C[\bar N,\bar N^e]$, 
`parameterized' by lapses and shifts that are vanishing on ${\cal 
S}$, is 

\begin{equation}
\Bigl\{C\bigl[N,N^a\bigr],C\bigl[\bar N,\bar N^a\bigr]\Bigr\}=C\bigl[
\bar N^eD_eN-N^eD_e\bar N,ND^a\bar N-\bar ND^aN-[N,\bar N]^a\bigr]. 
\label{eq:2.6}
\end{equation}
Furthermore, the new smearing fields $\bar N^eD_eN-N^eD_e\bar N$ and 
$ND^a\bar N-\bar ND^aN-[N,\bar N]^a$ are also vanishing on the 
boundary ${\cal S}$. Therefore, the constraints close to a Poisson 
algebra ${\cal C}$. 

Geometrically $N\vert_{\cal S}=0$, $N^a\vert_{\cal S}=0$ correspond 
to an evolution vector field $\xi^a=t^aN+N^a$ in the spacetime that 
is vanishing on ${\cal S}$, and hence the 1-parameter family of 
diffeomorphisms $\phi_t$ generated by $\xi^a$ leaves ${\cal S}$ fixed 
{\it pointwise}. This $\phi_t$ maps $\Sigma$ into a family $\Sigma_t$ 
of Cauchy surfaces for the {\it same} globally hyperbolic domain $D
(\Sigma)$ with the same boundary $\partial\Sigma_t={\cal S}$. 
According to Bergmann \cite{Be}, the gauge-invariant content of 
general relativity is the \emph{spacetime geometry}, and hence any two 
sets of information that specify the same spacetime geometry must be 
considered to be gauge equivalent. In particular, two Cauchy data sets 
determining the same globally hyperbolic domain are gauge equivalent 
in this sense. (For different interpretations see e.g. \cite{Kuchar, 
Ba}.) Therefore, the evolution vector fields $\xi^a$ that are 
vanishing at ${\cal S}$ are precisely the generators of \emph{gauge 
motions} of the actual quasi-local state in the \emph{spacetime}.

\subsubsection{The algebra of the basic Hamiltonians}
\label{sub-2.3.2}

If in the quasi-local Lagrangian phase space we choose the Lagrangian 
$L:=\frac{1}{2\kappa}\int_\Sigma(R-2\lambda+\chi_{ab}\chi^{ab}-
\chi^2)N\sqrt{\vert h\vert}{\rm d}^nx$, then for the basic Hamiltonian 
we obtain 

\begin{eqnarray}
H_0\bigl[K^e\bigr]:\!\!\!\!&=\!\!\!\!&C\bigl[K^e\bigr]+\int_\Sigma
 2D_a\Bigl(\tilde p{}^{ab}h_{bc}N^c\Bigr){\rm d}^nx= \nonumber \\
\!\!\!\!&=\!\!\!\!&C\bigl[K^e\bigr]-\frac{1}{\kappa}\oint_{\cal S}
K^a\Bigl(-v_at_bQ_c{}^{cb}+A_a\Bigr){\rm d}{\cal S}.
\label{eq:2.7}
\end{eqnarray}
Thus we already have a nontrivial a priori Hamiltonian boundary term, 
in which both $v_at_bQ_c{}^{cb}$ and $A_a$ depend on the actual choice 
for the normals $(t^a,v^a)$ of ${\cal S}$. This Hamiltonian can be 
made $2+(n-1)$-covariant if $K^e$ is restricted to be tangent to 
${\cal S}$ and, to cure the $SO(1,1)$-gauge dependence of the 
connection 1-form, if $\delta_eK^e=0$ is required. 

In fact \cite{szab06b}, evaluating the boundary terms in the total 
variation of $H_0$ we obtain that $H_0$ is functionally differentiable 
if the lapse is vanishing on ${\cal S}$, the shift is tangential to 
${\cal S}$, and the volume $(n-1)$-form is fixed on ${\cal S}$. 
However, this boundary condition is preserved by the evolution 
equations precisely when $\delta_eN^e=0$ is satisfied. With these 
boundary conditions the basic Hamiltonians form a Poisson algebra 
${\cal H}_0$, in which the quasi-local constraint algebra ${\cal C}$ 
is an ideal. By evaluating the basic Hamiltonian on the constraint 
surface we get a function on the algebra ${\cal H}_0/{\cal C}$ of 
observables, which provides a representation of the Lie algebra of 
the $\delta_e$-divergence-free vector fields on ${\cal S}$. Though 
the observable $O[N^e]:=-\frac{1}{\kappa}\oint_{\cal S}N^eA_e{\rm d}
{\cal S}$ behaves as angular momentum in certain special situations 
(see \cite{szab06b,szab07a}), but this can be non-zero even in 
Minkowski spacetime. This shows that the boundary term of $H_0$ 
should probably be present in the `ultimate' Hamiltonian, but still 
further terms are needed. 

Geometrically, $N\vert_{\cal S}=0$, $v_aN^a\vert_{\cal S}=0$ correspond 
to evolution vector fields \emph{tangential} to ${\cal S}$ on ${\cal 
S}$. The corresponding 1-parameter family of diffeomorphisms still maps 
$D(\Sigma)$ onto itself and preserves the boundary ${\cal S}$, but not 
pointwise. Its action on ${\cal S}$ preserves the volume $(n-1)$ form. 
Thus $H_0$ is certainly not the `ultimate' Hamiltonian, it is only an 
improved version of the constraint functions.

\subsubsection{On the differentiability of the `improved' basic 
Hamiltonian}
\label{sub-2.3.3}

The `bad' gauge dependence of the basic Hamiltonian (\ref{eq:2.7}) 
can be improved slightly by hand by adding $N\nu=K^at_av_bQ_c{}^{cb}$ 
to the integrand of the boundary integral. (Here $\nu:=\nu_{ab}q^{ab}$, 
the trace of the extrinsic curvature of ${\cal S}$ in $\Sigma$.) The 
resulting expression, 

\begin{equation}
H_1\bigl[K^e\bigr]:=C\bigl[K^e\bigr]-\frac{1}{\kappa}\oint_{\cal S}
K^a\Bigl({}^\bot\varepsilon_{ab}Q_c{}^{cb}+A_a\Bigr){\rm d}{\cal S},
\label{eq:2.8}
\end{equation}
has been derived in different forms by several authors \cite{HH,BLY, 
Anco} and used \cite{WY} to define quasi-local energy.\footnote{The 
same terms also appear in the general expression of the quasi-local 
quantities based on M\o ller's boost-gauge invariant, but still 
$O(1,n)$ gauge-dependent superpotential, and hence in the spinorial 
expressions based on the Nester--Witten 2-form in 3+1 dimensions, too. 
For details see the forthcoming updated version of \cite{szab04}.}
Here the first term in the boundary integral became $2+(n-1)$ 
covariant, but the second term still depends on the $SO(1,1)$ boost 
gauge. To cure this dependence, we still must require $\delta_eK^e=0$. 
In subsection \ref{sub-4.2} we will see that the boundary term of 
(\ref{eq:2.8}) emerges naturally among the boundary terms in the 
calculation of the Poisson bracket of two constraint functions (or 
Hamiltonians). Moreover, in that context there is a natural way of 
curing its $SO(1,1)$ boost-gauge dependence. On the other hand, as we 
will see, it does not seem to represent correctly the `composition 
law' of the lapses and shifts. 

Calculating the total variation of $H_1$, we can determine the 
condition of its functional differentiability. However, it is enough 
to calculate the total variation of the `correction term' $N\nu$ and 
to use the expression (4.2) of \cite{szab06b} for the total variation 
$\delta H_0[K^e]$ of the basic Hamiltonian. The total variation of 
$N\nu\sqrt{\vert q\vert}$ is 

\begin{eqnarray}
\delta\bigl(N\nu\sqrt{\vert q\vert}\bigr)\!\!\!\!&=\!\!\!\!&\Bigl(
 \nu\delta N+\frac{1}{2}Nv^e\bigl(D_e\delta h_{ab}\bigr)q^{ab}+
 \frac{1}{2}\nu Nq^{ab}\delta q_{ab}+ \nonumber \\
\!\!\!\!&+\!\!\!\!&\delta_a\bigl(N\delta v^a\bigr)-\bigl(\Delta_aN
 \bigr)\delta v^a-\nu Nv_a\delta v^a\Bigr)\sqrt{\vert q\vert}, 
 \nonumber
\end{eqnarray}
where we used decomposition (\ref{eq:2.3}). This, together with 
$\delta H_0[K^e]$, yields 

\begin{eqnarray}
\delta H_1\bigl[N,N^e\bigr]\!\!\!\!&=\!\!\!\!&C\bigl[\delta N,\delta 
 N^e\bigr]+\int_\Sigma\Bigl(\frac{\delta C[N,N^e]}{\delta h_{ab}}
 \delta h_{ab}+\frac{\delta C[N,N^e]}{\delta\tilde p^{ab}}\delta
 \tilde p^{ab}\Bigr){\rm d}^nx- \nonumber \\
-\frac{1}{\kappa}\oint_{\cal S}\Bigl\{\!\!\!\!&\nu\!\!\!\!&
 \delta N+\bigl(A_e-\tau v_e\bigr)\delta N^e+v_eN^eA_a\delta v^a
 -\tau v_eN^ev_a\delta v^a+\nonumber \\
\!\!\!\!&+\!\!\!\!&\frac{1}{2}\Bigl(-\nu^{ab}N+\tau^{ab}v_eN^e+ 
 \nonumber \\
\!\!\!\!&{}\!\!\!\!&+\bigl(\nu N-\tau v_eN^e+v^eD_eN+\chi_{cd}v^c
 v^cv_eN^e\bigr)q^{ab}\Bigr)\delta q_{ab}\Bigr\}{\rm d}{\cal S}. 
 \nonumber
\end{eqnarray}
Thus if the variations of the lapse and the shift were independent of 
the variations of the canonical variables, then the differentiability 
of $H_1[K^e]$ with respect to the metric $h_{ab}$ could be ensured by 
keeping the whole $n$-metric fixed on ${\cal S}$: $\delta h_{ab}\vert
_{\cal S}=0$. However, this condition is not $2+(n-1)$-covariant, and 
from the compatibility of this boundary condition with the evolution 
equation for $h_{ab}$ it follows that $K^e$ on ${\cal S}$ must satisfy 
$v^cv^d\nabla_{(c}K_{d)}=0$, $\Pi^c_av^d\nabla_{(c}K_{d)}=0$ and $\Pi
^c_a\Pi^d_b\nabla_{(c}K_{d)}=0$, which are conditions on the derivative 
of the lapse and shift in the direction $v^a$ normal to ${\cal S}$, 
too. Another possibility is that the induced metric $q_{ab}$ on ${\cal 
S}$ is fixed and the shift is tangent to ${\cal S}$. Apparently, this 
is a weaker condition for $(N,N^a)$ than what we had in subsection 
\ref{sub-2.3.2}. However, $v_eN^e\vert_{\cal S}=0$ is invariant with 
respect to the $SO(1,1)$ transformation of the normals $(t^a,v^a)$ 
only if we require $N\vert_{\cal S}=0$, too, i.e. we arrived back to 
the basic Hamiltonian. Hence, we conclude that \emph{the Hamiltonian 
(\ref{eq:2.8}) could be the `ultimate' quasi-local Hamiltonian only 
if the lapse and shift are not independent of the canonical variables}. 
Therefore, according to our expectation in subsection \ref{sub-2.1}, 
the lapse and the shift should satisfy a certain linear differential 
equation on ${\cal S}$. In subsection \ref{sub-4.1} we derive such an 
equation, but to motivate those investigations first we study the 
quasi-local canonical formulation of a single real scalar field.


\section{Illustration: Matter fields}
\label{sec-3}

\subsection{Conserved quantities and flux integrals for general matter 
fields}
\label{sub-3.1}

Let the matter fields be described by the symmetric energy-momentum 
tensor $T_{ab}$, which is divergence free if the field equations are 
satisfied. Let $K^e$ be any vector field, $\Sigma$ a smooth compact 
spacelike hypersurface with smooth boundary ${\cal S}:=\partial\Sigma$, 
and let us form the integral 

\begin{equation}
{\tt Q}_\Sigma\bigl[K^e\bigr]:=\int_\Sigma K_aT^{ab}\frac{1}{n!}
\varepsilon_{be_1...e_n}. \label{eq:3.1}
\end{equation}
Let $\xi^e$ be another, arbitrary smooth vector field on $M$, define 
$\Sigma_t$ to be the 1-parameter family of hypersurfaces by Lie 
dragging $\Sigma$ along the integral curves of $\xi^e$ such that 
$\Sigma_0=\Sigma$ and form the 1-parameter family of integrals 
(\ref{eq:3.1}) on these hypersurfaces. Then the derivative of these 
integrals with respect to the natural parameter $t$ along the 
integral curves of $\xi^e$ at $t=0$ is 

\begin{eqnarray}
\frac{\rm d}{{\rm d}t}{\tt Q}_\Sigma\bigl[K^e\bigr]=\int_\Sigma L_{\xi}
 \bigl(K_aT^{ab}\frac{1}{n!}\varepsilon_{be_1...e_n}\bigr)\!\!\!\!&=
 \!\!\!\!&\int_\Sigma\Bigl(T^{ab}\nabla_{(a}K_{b)}+\bigl(\nabla_aT
 ^{ab}\bigr)K_b\Bigr)\xi^ct_c{\rm d}\Sigma+  \nonumber \\
\!\!\!\!&+\!\!\!\!&\int_\Sigma\nabla_a\Bigl(\xi^aT^{bc}K_c-\xi^b
 T^{ac}K_c\Bigr)\frac{1}{n!}\varepsilon_{be_1...e_n}. \nonumber
\end{eqnarray}
However, the integrand of the second integral on the right can be 
rewritten into the exact $n$-form $-\frac{1}{(n-1)!}\nabla_{[e_1}
(\varepsilon_{e_2...e_{n-1}]ab}\xi^aT^{bc}K_c)$, and hence by the 
Stokes theorem it can be converted into a boundary integral. Thus 
finally we have 

\begin{equation}
\frac{\rm d}{{\rm d}t}{\tt Q}_\Sigma\bigl[K^e\bigr]=\int_\Sigma
\Bigl(T^{ab}\nabla_{(a}K_{b)}+\bigl(\nabla_aT^{ab}\bigr)K_b\Bigr)
\xi^ct_c{\rm d}\Sigma+\oint_{\cal S}\xi^a{}^\bot\varepsilon_{ab}T^{bc}
 K_c {\rm d}{\cal S}. \label{eq:3.2}
\end{equation}
Therefore, if the energy-momentum tensor is divergence free, there are 
two roots of the non-conservation of the quantity ${\tt Q}_\Sigma[K^e]$: 
The non-Killing nature of the vector field $K^a$ and the boundary 
integral. 

If $K^a$ is a Killing field, then the vanishing of the right hand side 
of (\ref{eq:3.2}) can be expected only for $\xi^a$ tangent to ${\cal 
S}$. The hypersurfaces $\Sigma_t$ that such a $\xi^a$ generates are 
such that the boundaries of all these coincide, $\partial\Sigma_t={\cal 
S}$, and they are simply other Cauchy surfaces for the same globally 
hyperbolic domain $D(\Sigma)$. Therefore, the ${\tt Q}_\Sigma[K^e]$ 
for Killing $K^e$ must in fact be conserved for $\xi^e$ tangential to 
${\cal S}$ on ${\cal S}$. For example, if $(M,g_{ab})$ is the 
Minkowski spacetime with Cartesian coordinates $\{x^{\ua}\}$, ${\ua}
=0,\dots,n$, then the general Killing field has the form $K_e=T_{\ua}
\nabla_ex^{\ua}+M_{\ua\ub}(x^{\ua}\nabla_ex^{\ub}-x^{\ub}\nabla_ex
^{\ua})$ for some constants $T_{\ua}$ and $M_{\ua\ub}=M_{[\ua\ub]}$. 
Then the coefficients of these constants in ${\tt Q}_\Sigma[K^e]=:
T_{\ua}{\tt P}^{\ua}+M_{\ua\ub}{\tt J}^{\ua\ub}$ define the 
quasi-local energy-momentum and angular momentum of the matter fields. 
These are conserved during the evolution with \emph{any} $\xi^e$ being 
tangent to ${\cal S}$ at ${\cal S}$, and transform in the correct, 
expected way under Poincar\'e transformations of the Cartesian 
coordinates. Hence the quasi-local quantities can be thought of as 
being associated with ${\cal S}$ or with the whole Cauchy development 
$D(\Sigma)$ of $\Sigma$. 

To understand the meaning of the boundary integral in (\ref{eq:3.2}), 
suppose that $\xi^e$ is not tangent to ${\cal S}$. Since the area 
2-form ${}^\bot\varepsilon_{ab}$ annihilates the part of $\xi^e$ 
tangential to ${\cal S}$, without loss of generality we may assume 
that $\xi^e$ is orthogonal to ${\cal S}$. Let $B$ denote the union 
of the boundaries $\partial\Sigma_t$ for all $\vert t\vert<\epsilon$ 
for some positive $\epsilon$, i.e. the hypersurface that ${\cal S}$ 
sweeps out by Lie dragging along the integral curves of $\xi^e$. 
(This $B$ is a smooth, regular hypersurface only if $\xi^e$ is 
nowhere vanishing on ${\cal S}$. At the zeros of $\xi^e$ the 
boundaries $\partial\Sigma_t$ intersect each other, and at these 
points $B$ collapses to $(n-1)$ dimensional.) Then, by construction, 
$\xi^a{}^\bot\varepsilon_{ab}$ is a (not normalized) normal 1-form 
to the (regular parts of) $B$ on ${\cal S}$. Thus, the integrand of 
the boundary integral on the right hand side of (\ref{eq:3.2}) is 
the flux density of the current $T^{ab}K_b$ through $B$ at ${\cal S}$ 
weighted by the `length' of $\xi^e$. Therefore, for small enough 
$\Delta t$, $\oint_{\cal S}\xi^a{}^\bot\varepsilon_{ab}T^{bc}K_c{\rm 
d}{\cal S}\Delta t$ is the \emph{flux of the current $T^{ab}K_b$ 
through $B$ between $\partial\Sigma_0$ and $\partial\Sigma_{\Delta 
t}$}. The root of the non-conservation of the quasi-local quantities 
${\tt Q}_\Sigma[K^e]$ even for Killing $K^e$ is that the actual 
system, surrounded at a given instant by ${\cal S}$, is \emph{not} 
closed, and there can be non-trivial incoming and outgoing flows of 
energy-momentum and angular momentum. 

In particular, if $K^e$ is the time translational Killing field, $K_e
=\nabla_ex^0$, then ${\tt Q}_\Sigma[K^e]={\tt P}^0$, the quasi-local 
energy. If $T^{ab}$ satisfies the dominant energy condition, then this 
is non-negative (and zero if and only if $T^{ab}$ is vanishing on $D
(\Sigma)$). Then by (\ref{eq:3.2}) $\frac{\rm d}{{\rm d}t}{\tt P}^0$ 
is non-negative for outward pointing (i.e. for which $v_e\xi^e<0$) 
spacelike or null $\xi^e$, non-positive for inward pointing (i.e. 
$v_e\xi^e>0$) spacelike or null $\xi^e$, and does not have a definite 
sign for timelike $\xi^e$. In the first case only the incoming energy 
flow can cross $B$, yielding energy gain; in the second only the 
outgoing energy flow can cross $B$, yielding energy loss; while in 
the case of timelike $B$, both incoming and outgoing energy flows may 
be present. 

As a conclusion, first, a distinction between evolution vector fields 
$\xi^e$ generating the (e.g. time) evolution of the state and the 
generators of the quasi-local (conserved) quantities $K^e$ must be 
made. (This view was already adopted already both for matter fields in 
\cite{szab04,szab05} and for gravitational fields that are 
asymptotically flat at spatial infinity in \cite{szab03,szab06a}.) 
Moreover, the boundary integral appearing in the `time' derivative of 
the `conserved' quantities describes the flux of the incoming and 
outgoing energy-momentum and angular momentum flows. We will see in 
subsection \ref{sub-3.2.2} that a detailed and systematic quasi-local 
Hamiltonian analysis of a single real scalar field exactly reproduces 
the result (\ref{eq:3.2}), where the boundary integral emerges as the 
boundary term in the Poisson bracket of two Hamiltonians. Though the 
gravitational `field' does not have any well-defined energy-momentum 
density but admits a Hamiltonian formulation, an analogous result may 
be expected for the gravitational `field' as well: the Poisson 
boundary term must have a physical meaning (and hence must be gauge 
invariant) for appropriately defined `quasi-symmetry generators' $K^e$ 
on ${\cal S}$.


\subsection{Quasi-local Hamiltonian description of the scalar field}
\label{sub-3.2}

\subsubsection{The quasi-local phase space and the Hamiltonian}
\label{sub-3.2.1}

Let $\Phi$ be a real scalar field on $M$, whose dynamics in the 
spacetime is governed by the Lagrangian ${\cal L}:=\frac{1}{2}g^{ab}
(\nabla_a\Phi)(\nabla_b\Phi)-V$, where the potential $V=V(\Phi)$ is 
a given \emph{algebraic} function of the scalar field. For the sake 
of concreteness, we may assume that this has the form $V=\frac{1}{2}
m^2\Phi^2+\frac{1}{4}\mu\Phi^4$, i.e. the scalar field is of 
rest-mass $m$ and $\mu$ is its self-interaction parameter. The 
covariant field equation and the energy-momentum tensor, respectively, 
are 

\begin{eqnarray}
&{}&\nabla_a\nabla^a\Phi+\frac{\partial V}{\partial\Phi}=0, 
 \label{eq:3.3.a} \\
&{}&T_{ab}=\bigl(\nabla_a\Phi\bigr)\bigl(\nabla_b\Phi\bigr)-\frac{1}
 {2}g_{ab}\bigl(\nabla_e\Phi\bigr)\bigl(\nabla^e\Phi\bigr)+g_{ab}V. 
 \label{eq:3.3.b}
\end{eqnarray}
$T_{ab}$ with the explicit form of the potential $V$ above satisfies 
the dominant energy condition precisely when $\mu\geq0$. 

The basis of the quasi-local canonical formulation of the theory of 
scalar field is the quasi-local configuration space ${\cal Q}(\Sigma)$, 
the space of the smooth real scalar fields on $\Sigma$ satisfying 
certain, not-yet-specified boundary conditions. The quasi-local 
Hamiltonian phase space is its `cotangent bundle' $T^*{\cal Q}
(\Sigma)$, endowed with its natural symplectic structure. The canonical 
momenta are scalar densities $\tilde\Pi$ on $\Sigma$. Using the 
Lagrangian $L:T{\cal Q}(\Sigma)\rightarrow\mathbb{R}$, defined by 
$L:=\int_\Sigma{\cal L}N\sqrt{\vert h\vert}{\rm d}^nx$ and considered 
to be the function of the Lagrange variables $(\Phi,\dot\Phi)$, the 
standard canonical formalism yields for the momenta that $\tilde\Pi=
t^a(\nabla_a\Phi)\sqrt{\vert h\vert}=\frac{1}{N}(\dot\Phi-N^eD_e\Phi)
\sqrt{\vert h\vert}$ and for the Hamiltonian $H:T^*{\cal Q}(\Sigma)
\rightarrow\mathbb{R}$, introduced by $H:=\int_\Sigma(\tilde\Pi\dot
\Phi-{\cal L}N\sqrt{\vert h\vert}){\rm d}^nx$, that 

\begin{equation}
H=\int_\Sigma\Bigl\{N\Bigl(\frac{1}{2}\frac{\tilde\Pi^2}{\vert h
\vert}-\frac{1}{2}h^{ab}\bigl(D_a\Phi\bigr)\bigl(D_b\Phi\bigr)+V\Bigr)
\sqrt{\vert h\vert}+N^e\bigl(D_e\Phi\bigr)\tilde\Pi\Bigr\}{\rm d}^nx. 
\label{eq:3.4}
\end{equation}
A straightforward calculation shows that the coefficient of $N$ is 
just the energy density part $\mu:=T_{ab}t^at^b$ and the coefficient 
of $N^e$ is just the momentum density part $J_e:=P^a_eT_{ab}t^b$ of 
the symmetric energy-momentum tensor. Thus we can also write $H=\int
_\Sigma K^aT_{ab}t^b{\rm d}\Sigma$, where $K^a=Nt^a+N^a$. If more 
than one Hamiltonians `parameterized' by different lapse-shift pairs 
are considered, then to indicate which lapse-shift is used we write 
the Hamiltonian as $H[N,N^e]$ or $H[K^e]$. 

Since $H$ depends on the `parameters' $N$ and $N^e$, the spatial 
metric $h_{ab}$ and the momentum variable $\tilde\Pi$ 
\emph{algebraically}, $H$ is functionally differentiable with respect 
to them, independently of the boundary conditions on ${\cal S}$. 
(Though in the present context the functional differentiability with 
respect to $N$, $N^e$ and $h_{ab}$ does not have any significance, in 
subsection \ref{sub-4.1.5}, where we consider the Einstein--scalar 
system, $h_{ab}$ will be the gravitational configuration variable, and 
hence the functional differentiability with respect to $h_{ab}$ will 
be important.) The corresponding functional derivatives themselves are 

\begin{equation}
\frac{\delta H}{\delta N}=\mu\sqrt{\vert h\vert}, \hskip 10pt
\frac{\delta H}{\delta N^a}=J_a\sqrt{\vert h\vert}, \hskip 10pt
\frac{\delta H}{\delta h_{ab}}=\frac{1}{2}N\sigma^{ab}\sqrt{\vert 
 h\vert}, \label{eq:3.5}
\end{equation}
where $\sigma_{ab}:=T_{cd}P^c_aP^d_b$, the spatial stress part of the 
symmetric energy-mom\-entum tensor (\ref{eq:3.3.b}), and 

\begin{equation}
\frac{\delta H}{\delta\tilde\Pi}=N\frac{\tilde\Pi}{\sqrt{\vert h
\vert}}+N^eD_e\Phi. 
 \label{eq:3.6}
\end{equation}
Nevertheless, the condition of the functional differentiability with 
respect to $\Phi$ is 

\begin{equation}
\oint_{\cal S}v^a\Bigl(N_a\Pi-N\bigl(D_a\Phi\bigr)\Bigr)\delta\Phi
{\rm d}{\cal S}=0. 
\label{eq:3.7}
\end{equation}
(Here $\Pi$ is the `de-densitized' canonical momentum: $\tilde\Pi=:
\Pi\sqrt{\vert h\vert}$.) (\ref{eq:3.7}) can be ensured either by 
fixing the configuration variable $\Phi$ on ${\cal S}$, $\delta\Phi
\vert_{\cal S}=0$, or by requiring the vanishing of the coefficient 
of $\delta\Phi$ in (\ref{eq:3.7}). Under any of these conditions $H$ 
is functionally differentiable, and its functional derivative with 
respect to the configuration variable is 

\begin{equation}
\frac{\delta H}{\delta\Phi}=-D_a\Bigl(N^a\tilde\Pi-N\bigl(D^a\Phi
\bigr)\sqrt{\vert h\vert}\Bigr)+N\frac{\partial V}{\partial\Phi}
\sqrt{\vert h\vert}. \label{eq:3.8} 
\end{equation}
Then the canonical equations of motion are 

\begin{eqnarray}
&{}&\dot\Phi=\frac{\delta H}{\delta\tilde\Pi}=N\Pi+N^aD_a\Phi, 
\label{eq:3.9.a} \\
&{}&\dot{\tilde\Pi}=-\frac{\delta H}{\delta\Phi}=-D_a\Bigl(N\sqrt{
\vert h\vert}D^a\Phi-N^a\tilde\Pi\Bigr)-N\frac{\partial V}{\partial
\Phi}\sqrt{\vert h\vert}. \label{eq:3.9.b}
\end{eqnarray}
The first is equivalent to the definition of the time derivative of 
$\Phi$, while the second to the field equation (\ref{eq:3.3.a}). 

Returning to the boundary conditions, the first, $\delta\Phi\vert
_{\cal S}=0$, is analogous to the Dirichlet boundary condition in 
electrostatics. However, according to our requirement (ii) in subsection 
\ref{sub-2.1.1}, the evolution equation (\ref{eq:3.9.a}) must preserve 
this, and hence we obtain that at ${\cal S}$ the canonical variables 
must satisfy 

\begin{equation}
N\Pi+N^aD_a\Phi=0. \label{eq:3.10}
\end{equation}
Thus, the canonical momentum (weighted by the lapse) must be linked 
to the derivative of $\Phi$ in the direction of the shift. In 
particular, for vanishing shift $\tilde\Pi$ must be vanishing on 
${\cal S}$. 
The other possible boundary condition coming from (\ref{eq:3.7}) is 
the requirement of the vanishing of the coefficient of $\delta\Phi$ 
on ${\cal S}$: 

\begin{equation}
v^aN_a\Pi-Nv^aD_a\Phi=0. \label{eq:3.11}
\end{equation}
Thus, the normal directional derivative of $\Phi$ (weighted by the 
lapse) is linked to the canonical momentum on ${\cal S}$. This is 
analogous to the (generalized) Neumann-type boundary condition in 
electrostatics. (The Dirichlet- and Neumann-type boundary conditions 
appear naturally in a covariant phase-space context too. For details, 
see \cite{AT1,AT2}.)

\subsubsection{The Poisson boundary term and the flux integral}
\label{sub-3.2.2}

Let $(N,N^e)$ and $(\bar N,\bar N^e)$ be two lapse-shift pairs, assume 
that both $H[N,N^a]$ and $H[\bar N,\bar N^a]$ are differentiable and 
let us calculate the Poisson bracket of two Hamiltonians parameterized 
by them. By integration by parts, a direct calculation yields that it is 

\begin{eqnarray}
&{}&\Bigl\{H\bigl[N,N^e\bigr],H\bigl[\bar N,\bar N^e\bigr]
 \Bigr\}= \label{eq:3.12}\\
&{}&=H\Bigl[\bar N^aD_aN-N^aD_a\bar N,ND^a\bar N-\bar ND^aN-
 \bigl[N,\bar N\bigr]^a\Bigr]+\nonumber \\ 
&{}&+\int_\Sigma\Bigl(ND_{(a}\bar N_{b)}-\bar ND_{(a}N_{b)}
 \Bigr)\sigma^{ab}{\rm d}\Sigma- \nonumber \\
&{}&-\oint_{\cal S}v_a\Bigl\{\bigl(N^a\bar N^b-\bar N^a
 N^b\bigr)J_b+\bigl(\bar NN_b-N\bar N_b\bigr)\sigma^{ab}+\bigl(\bar N
 N^a-N\bar N^a\bigr)\Pi^2\Bigr\}{\rm d}
 {\cal S}. \nonumber
\end{eqnarray}
Note that the new lapse and shift that parameterize the Hamiltonian on 
the right are exactly those that appeared in the constraint algebra 
of Einstein's theory (see subsection \ref{sub-2.3.1}). But in addition 
to the Hamiltonian $H$ the spatial integral of the spatial stress, 
contracted with the Killing operators acting on the shift vectors (and 
weighted by the lapses), also appears. These operators can be replaced 
by the \emph{spacetime} Killing operators acting on appropriately 
defined \emph{spacetime} vector fields. Indeed, let us fix a foliation 
of the spacetime with lapse $M$ and a compatible evolution vector field 
$\xi^e:=Mt^e+M^e$, where $t^e$ is the future pointing unit timelike 
normal of the leaves of this foliation, and define $K^a:=Nt^a+N^a$ and 
$\bar K^a:=\bar Nt^a+\bar N^a$. Then the complete $n+1$ decomposition 
of the Killing operator $\nabla_{(a}\bar K_{b)}$ with respect to this 
foliation and evolution vector field $\xi^e$ is 

\begin{eqnarray}
Mt^ct^d\nabla_{(c}\bar K_{d)}\!\!\!\!&=\!\!\!\!&\dot{\bar N}+\bar 
 N^aD_aM-M^aD_a\bar N, \label{eq:3.13.a} \\
2Mh^{ac}t^d\nabla_{(c}\bar K_{d)}\!\!\!\!&=\!\!\!\!&\dot{\bar N}^a+
 MD^a\bar N-\bar ND^aM-\bigl[M,\bar N\bigr]^a, \label{eq:3.13.b} \\
P^c_aP^d_b\nabla_{(c}\bar K_{d)}\!\!\!\!&=\!\!\!\!&\bar N\chi_{ab}+
 D_{(a}\bar N_{b)}, \label{eq:3.13.c} 
\end{eqnarray}
where $\dot{\bar N}$ and $\dot{\bar N}^a$ denote the time derivative 
of $\bar N$ and $\bar N^a$ with respect to $\xi^e$, respectively, 
introduced in subsection \ref{sub-2.2.1}. Note that while the 
normal-normal and normal-tangential parts of the spacetime Killing 
operator depend on $M$ and $M^a$, its spatial projection does not. It 
is well defined even on a single spacelike hypersurface. Thus, by 
(\ref{eq:3.13.c}), the integrand of the $n$-dimensional integral on the 
right-hand side of (\ref{eq:3.12}) is $\sigma^{ab}(N\nabla_{(a}\bar 
K_{b)}-\bar N\nabla_{(a}K_{b)})$. 

On the other hand, contrary to the bulk terms, the boundary integral 
(which we call the Poisson boundary term) apparently contains the 
canonical momentum explicitly, and not only through the various parts 
of the symmetric energy momentum tensor. However, if we take into 
account any of the boundary conditions (\ref{eq:3.10}) and 
(\ref{eq:3.11}), then the Poisson boundary term can be rewritten 
purely in terms of the energy-momentum tensor. In fact, if 
(\ref{eq:3.10}) holds (both for $(N,N^a)$ and $(\bar N,\bar N^a)$), 
then by $\bar N\Pi^2=-\bar N^a(D_a\Phi)\Pi=-\bar N^aJ_a$ the first 
and the third terms in the boundary integral of (\ref{eq:3.12}) cancel 
each other, and there remains only $v_a\sigma^{ab}(\bar NN_b-N\bar 
N_b)$. Similarly, using the explicit form of the spatial stress 
$\sigma_{ab}$, we can write $\sigma^{ab}\bar NN_b=\bar NN^a\mu+N^a
\bar N^bJ_b-N\bar NJ^a$. Then, however, it is straightforward to 
check that the integrand of the boundary integral is an expression 
of the energy-momentum tensor, and the final form of (\ref{eq:3.12}) 
is 

\begin{eqnarray}
&{}&\Bigl\{H\bigl[N,N^e\bigr],H\bigl[\bar N,\bar N^e\bigr]\Bigr\}=
 \nonumber \\
&{}&=H\Bigl[\bar N^aD_aN-N^aD_a\bar N,ND^a\bar N-\bar ND^aN-\bigl[N,
 \bar N\bigr]^a\Bigr]+\nonumber \\ 
&{}&+\int_\Sigma\Bigl(N\nabla_{(a}\bar K_{b)}-\bar N\nabla_{(a}K_{b)}
 \Bigr)\sigma^{ab}{\rm d}\Sigma+\oint_{\cal S}K^a{}^\bot\varepsilon
 _{ac}T^c{}_b\bar K^b{\rm d}{\cal S}. \label{eq:3.14}
\end{eqnarray}
Using the boundary condition it is easy to check that the boundary 
integral is anti-symmetric in $K^a\bar K^b$, as it should be because 
every other term in (\ref{eq:3.14}) changes sign if we interchange 
$(N,N^a)$ and $(\bar N,\bar N^a)$. 
Similarly, if (\ref{eq:3.11}) holds, then the last term in the 
boundary integral in (\ref{eq:3.12}) is vanishing, and $v_a\sigma
^{ab}\bar NN_b$ $=v_aN^a\bar N\mu+v_a\bar N^aN^bJ_b-\bar NNv^bJ_b$ 
holds. Using this expression, (\ref{eq:3.12}) again takes the form 
(\ref{eq:3.14}). 

Recalling that in the spacetime picture the Hamiltonian $H[N,N^a]$ 
is the flux integral of the Lorentz-covariant current $T^{ab}K_b$, 
it is natural to ask for the spacetime-covariant form of 
(\ref{eq:3.14}), too. To derive this, let us decompose the Lie 
bracket $[K,\bar K]^a$ of the spacetime vector fields $K^a$ and 
$\bar K^a$ with respect to the foliation and evolution vector field 
above. It is 

\begin{eqnarray}
\bigl[K,\bar K\bigr]^at_a\!\!\!\!&=\!\!\!\!&t^bt^c\Bigl(N\nabla_{(b}
 \bar K_{c)}-\bar N\nabla_{(b}K_{c)}\Bigr)+N^aD_a\bar N-\bar N^aD_a
 N, \label{eq:3.15.a} \\
\bigl[K,\bar K\bigr]^bP^a_b\!\!\!\!&=\!\!\!\!&2h^{ab}t^c\Bigl(N\nabla
 _{(b}\bar K_{c)}-\bar N\nabla_{(b}K_{c)}\Bigr)- \nonumber \\
&{}&-\bigl(ND^a\bar N-\bar ND^aN\bigr)+\bigl[N,\bar N\bigr]^a. 
 \label{eq:3.15.b}
\end{eqnarray}
Thus \emph{the new lapse and shift both in (\ref{eq:2.6}) and 
(\ref{eq:3.12}) appear as the lapse and shift parts of the Lie bracket 
of the spacetime vector fields up to the spacetime Killing operators}. 
Substituting these into (\ref{eq:3.14}) and using the notation $H[K^e]
=H[N,N^e]$, we obtain 

\begin{eqnarray}
\Bigl\{H\bigl[K^e\bigr],H\bigl[\bar K^e\bigr]\Bigr\}\!\!\!\!&=
 \!\!\!\!&-H\Bigl[\bigl[K,\bar K\bigr]^e\Bigr]+\int_\Sigma t^c\Bigl(
 K_c\nabla_{(a}\bar K_{b)}-\bar K_c\nabla_{(a}K_{b)}\Bigr)T^{ab}{\rm 
 d}\Sigma+ \nonumber \\
\!\!\!\!&+\!\!\!\!&\frac{1}{2}\oint_{\cal S}\bigl(K^a\bar K^b-\bar 
 K^aK^b\bigr)\,{}^\bot\varepsilon_{ac}T^c{}_b{\rm d}{\cal S}. 
 \label{eq:3.16}
\end{eqnarray}
Therefore, the quasi-local Hamiltonians of the real scalar field do 
\emph{not} form a Poisson algebra. The two roots of this failure are 
the non-Killing nature of the vector fields $K^a$ and $\bar K^a$ and 
the boundary integral. The latter is precisely the boundary integral 
of (\ref{eq:3.2}). Our aim is to recover the whole of (\ref{eq:3.2}).

To do this, let us calculate the time derivative of $H[\bar N,\bar 
N^a]$ with respect to $\xi^a$ in the spacetime. It is 

\begin{eqnarray}
\frac{\rm d}{{\rm d}t}H\bigl[\bar N,\bar N^a]\!\!\!\!&=\!\!\!\!&
 \int_\Sigma\Bigl(\mu\dot{\bar N}+J_a\dot{\bar N}^a+\frac{1}{2}
 \bar N\sigma^{ab}\dot h_{ab}\Bigr){\rm d}\Sigma+ \nonumber \\
\!\!\!\!&+\!\!\!\!&\int_\Sigma\Bigl(\frac{\delta H[\bar N,\bar N
 ^a]}{\delta\Phi}\dot\Phi+\frac{\delta H[\bar N,\bar N^a]}{\delta
 \tilde\Pi}\dot{\tilde\Pi}\Bigr){\rm d}^nx= \nonumber \\
\!\!\!\!&=\!\!\!\!&\int_\Sigma\Bigl(\mu\dot{\bar N}+J_a\dot{\bar 
 N}^a+\bar N\sigma^{ab}\bigl(M\chi_{ab}+D_{(a}M_{b)}\bigr)\Bigr)
{\rm d}\Sigma+ \nonumber \\
\!\!\!\!&+\!\!\!\!&\Bigl\{H\bigl[M,M^a\bigr],H\bigl[\bar N,\bar 
N^a\bigr]\Bigr\}, \label{eq:3.17}
\end{eqnarray}
where in the first step we used the functional differentiability of 
$H[\bar N,\bar N^a]$, and in the second the canonical equations of 
motion with the Hamiltonian $H[M,M^a]$. Finally, by the expression 
(\ref{eq:3.14}) of the Poisson bracket and the projections 
(\ref{eq:3.13.a})-(\ref{eq:3.13.c}) of the spacetime Killing operator, 
we obtain 

\begin{equation}
\frac{\rm d}{{\rm d}t}H\bigl[\bar N,\bar N^a]=\int_\Sigma\xi^ct_c T
^{ab}\nabla_{(a}\bar K_{b)}{\rm d}\Sigma+\oint_{\cal S}\xi^a\,{}^\bot
\varepsilon_{ac}T^{cb}\bar K_b{\rm d}{\cal S}. \label{eq:3.18}
\end{equation}
Thus we recovered (\ref{eq:3.2}), whose boundary integral appeared 
here as the Poisson boundary term.


\section{The Poisson boundary terms in GR}
\label{sec-4}

\subsection{The Poisson boundary terms}
\label{sub-4.1}

\subsubsection{The formal Poisson brackets}
\label{sub-4.1.1}

Though in subsection \ref{sub-2.1.1} we defined the Poisson bracket 
for \emph{functionally differentiable} functions on the phase space, 
we can define the \emph{formal} Poisson bracket of any two constraint 
functions $C[N,N^e]$ and $C[\bar N,\bar N^e]$ by the integral of 
their \emph{formal} functional derivatives, independently of their 
functional differentiability. A lengthy but straightforward 
calculation gives (or see \cite{szab03, szab06b}) 

\begin{eqnarray}
\Bigl\{C\bigl[N,N^a\bigr]\!\!\!\!\!\!&,\!\!\!\!\!\!&C\bigl[\bar N,
 \bar N^a\bigr]\Bigr\}=C\Bigl[\bar N^eD_eN-N^eD_e\bar N\, ,\,ND^a
 \bar N-\bar ND^aN-\bigl[N,\bar N\bigr]^a\Bigr]- \nonumber \\
\!\!\!\!&-\!\!\!\!&\oint_{\cal S}2v_ap^{ab}\Bigl(ND_b\bar N-\bar ND_bN
 -\bigl[N,\bar N\bigr]_b\Bigr){\rm d}{\cal S}-\nonumber \\
\!\!\!\!&-\!\!\!\!&\oint_{\cal S}2p^{ab}\Bigl(v_eN^eD_a\bar N_b-v_e
 \bar N^eD_aN_b\Bigr){\rm d}{\cal S}-\nonumber \\
\!\!\!\!&-\!\!\!\!&\oint_{\cal S}\frac{1}{\kappa}\Bigl\{\frac{1}{2}
 \Bigl(R-2\lambda+\chi^2-\chi_{ab}\chi^{ab}\Bigr)\bigl(N\bar N^e-\bar 
 NN^e\bigr)v_e-v^aR_{ab}\bigl(N\bar N^b-\bar NN^b\bigr)\nonumber \\
&{}&+v^a\bigl(D_aN\bigr)\bigl(D_b\bar N^b\bigr)-\bigl(D^bN\bigr)
 \bigl(D_b\bar N_a\bigr)v^a-\nonumber \\
&{}&-v^a\bigl(D_a\bar N\bigr)\bigl(D_bN^b\bigr)+\bigl(D^b\bar N\bigr)
 \bigl(D_bN_a\bigr)v^a\Bigr\}{\rm d}{\cal S}. \label{eq:4.1}
\end{eqnarray}
A well-known highly non-trivial property of the constraint functions 
is that in their formal Poisson bracket, the genuine $n$-dimensional 
integral is also a constraint function (with the new lapse $\bar N^e
D_eN-N^eD_e\bar N$ and the new shift $ND^a\bar N-\bar ND^aN-\bigl[N,
\bar N\bigr]^a$), and the remaining terms are all boundary integrals. 
It might also be interesting to note that the first two terms on the 
right together is just the basic Hamiltonian of subsection 
\ref{sub-2.3.2} parameterized by the new lapse and shift.

\subsubsection{The main idea}
\label{sub-4.1.2}

We learnt in subsection \ref{sub-3.2.2} that the Poisson boundary term 
in $\{H[\xi^a],H[\bar K^a]\}$ describes, at least for appropriately 
chosen Killing fields $\bar K^a$, the infinitesimal flux of 
energy-momentum and angular momentum flows through the hypersurface 
that is generated by Lie dragging ${\cal S}$ along the integral 
curves of $\xi^a$. However, if we could expect that the Poisson 
boundary term has a physical meaning in general relativity too, then 
(in addition to the requirement of the functional differentiability 
of the Hamiltonian and the compatibility of the boundary conditions 
and the evolution equations) we would have a further condition that 
we could use to find the `ultimate' Hamiltonian boundary term and 
the boundary conditions. 

Thus suppose for a moment that we already have the `ultimate' 
Hamiltonian boundary term and the boundary conditions both for the 
canonical variables and the lapse and the shift; and hence $H[N,N^a]$, 
given by (\ref{eq:1.3}), is functionally differentiable. Then the 
Poisson bracket of two such Hamiltonians is precisely the \emph{formal} 
Poisson bracket of the two \emph{constraints} with the same lapses and 
shifts, which is already given explicitly by (\ref{eq:4.1}). However, 
if our expectation is correct, then this Poisson bracket must be the 
sum of the `correct' Hamiltonian (parameterized by the new lapse and 
shift) and another physical quantity (being analogous to the 
energy-momentum and angular momentum fluxes). Consequently, since both 
are gauge invariant, the Poisson boundary term \emph{must} also be 
gauge invariant. In particular, this Poisson boundary term 

\begin{description}

\item{1.} should depend only on $N\vert_{\cal S}$, $\bar N\vert_{\cal 
  S}$, $N^a\vert_{\cal S}$ and $\bar N^a\vert_{\cal S}$, but not on 
  their normal derivatives, e.g. on $v^eD_eN\vert_{\cal S}$ or $v^e
  D_eN^a\vert_{\cal S}$; 

\item{2.} should depend only on the geometry of ${\cal S}$, but not on 
  the geometry of $\Sigma$ at its boundary ${\cal S}$; 

\item{3.} must be $2+(n-1)$-covariant and, in particular, it must be 
  independent of the actual choice for the normals $(t^a,v^a)$ to 
  ${\cal S}$. 
\end{description}
Therefore, we must check whether or not the boundary terms in 
(\ref{eq:4.1}) satisfy these criteria. This will be done by rewriting 
it in a form adapted to ${\cal S}$, using the ideas and notions 
summarized in subsection \ref{sub-2.2.2}.

\subsubsection{The covariant form of the Poisson boundary term}
\label{sub-4.1.3}

Using the definitions of subsection \ref{sub-2.2.2}, by a systematic 
decomposition of every tensor field and derivative operator according 
to $P^a_b=\Pi^a_b-v^av_b$ we rewrite the integrand of every boundary 
integral in (\ref{eq:4.1}). Our ultimate aim is to obtain a 
$2+(n-1)$-covariant form and, in particular, in terms of the 
spacetime vector fields $K^e:=Nt^e+N^e$ and $\bar K^e:=\bar Nt^e+\bar 
N^e$ rather than the individual lapses and shifts. 

First, by a tedious but straightforward computation for the integrand 
$I_1$ of the first boundary integral in (\ref{eq:4.1}), we obtain 

\begin{eqnarray}
2\kappa I_1\!\!\!\!&=\!\!\!\!&\Bigl(K^e\delta_e\bar K^a-\bar K^e
 \delta_eK^a\Bigr)\bigl(\tau v_a-A_a\bigr)+ \nonumber \\
\!\!\!\!&+\!\!\!\!&\Bigl(\tau t_cA_d-A^b\bigl(\tau_{bc}t_d-\nu_{bc}
 v_d\bigr)\Bigr)\bigl(K^c\bar K^d-\bar K^cK^d\bigr)- \nonumber \\
\!\!\!\!&-\!\!\!\!&A_bA^b{}^\bot\varepsilon_{cd}K^c\bar K^d+ 
 \nonumber \\
\!\!\!\!&+\!\!\!\!&\Bigl(\bigl(\delta_e\bar K^a\bigr)t_at_bK^b-
 \bigl(\delta_eK^a\bigr)t_at_b\bar K^b\Bigr)A^e+ \nonumber \\
\!\!\!\!&+\!\!\!\!&\Bigl(\bar Nv^eD_eN-Nv^eD_e\bar N\Bigr)\tau+ 
 \nonumber \\
\!\!\!\!&+\!\!\!\!&\Bigl(v_eN^ev^aD_a\bar N_b-v_e\bar N^ev^aD_aN_b
 \Bigr)\bigl(A^b-\tau v^b\bigr). \label{eq:4.2.a}
\end{eqnarray}
(To reproduce this formula (and the similar ones below), it seems 
useful to calculate first the projections of various quantities, e.g. 
$2\kappa v_av_bp^{ab}=\chi_{ab}v^av^b+\chi=\tau_{ab}q^{ab}=\tau$, 
$2\kappa v_ap^{ab}q_{bc}=A_c$ or $2\kappa p^{cd}q_{ca}q_{db}=\tau
_{ab}-q_{ab}(\tau-\chi_{cd}v^cv^d)$. Note that $\chi_{ab}v^av^b$ 
cannot be expressed by quantities defined only on ${\cal S}$.) 

Similarly, the integrand $I_2$ of the second boundary integral is 

\begin{eqnarray}
2\kappa I_2\!\!\!\!&=\!\!\!\!&\Bigl(v_eK^e\Delta_a\bar K_b-v_e\bar 
 K^e\Delta_aK_b\Bigr)\bigl(\tau^{ab}-\tau q^{ab}\bigr)+\nonumber \\
\!\!\!\!&+\!\!\!\!&\Bigl(v_eK^e\bigl(\Delta_a\bar K^a\bigr)-v_e
 \bar K^e\bigl(\Delta_aK^a\bigr)\Bigr)\chi_{cd}v^cv^d+\nonumber \\
\!\!\!\!&+\!\!\!\!&\Bigl(\tau_{ab}\tau^{ab}-\tau^2+\tau\chi_{ab}
 v^av^b-A_bA^b\Bigr)\,{}^\bot\varepsilon_{cd}K^c\bar K^d+ \nonumber \\
\!\!\!\!&+\!\!\!\!&\Bigl(v_e\bar K^e\bigl(\Delta_aK_b\bigr)v^b-
 v_eK^e\bigl(\Delta_a\bar K_b\bigr)v^b\Bigr)A^a+ \nonumber \\
\!\!\!\!&+\!\!\!\!&\Bigl(v_eK^ev^aD_a\bar N_b-v_e\bar K^ev^aD_aN_b
 \Bigr)\bigl(\tau v^b-A^b\bigr). \label{eq:4.2.b}
\end{eqnarray}
Before rewriting the integrand $I_3$ of the third boundary integral 
let us observe that the Ricci tensor and the curvature scalar appear 
in $I_3$ just like in the `constraint parts' $v^av^bG_{ab}$ and $v^a
G_{ab}\Pi^b_c$ of the Einstein tensor of the $n$-dimensional intrinsic 
(spatial) geometry $(\Sigma,h_{ab})$. Expressing these in terms of 
the curvature scalar ${}^{\cal S}R$ and the extrinsic curvature $\nu
_{ab}$ of ${\cal S}$ (and its derivative $\delta_e\nu_{ab}$), a direct 
but quite lengthy calculation gives that 

\begin{eqnarray}
2\kappa I_3\!\!\!\!&=\!\!\!\!&-\lambda{}^\bot\varepsilon_{cd}K^c
 \bar K^d+\delta_a\Bigl(\bigl(\nu^{ab}-\nu q^{ab}\bigr)\bigl(N\bar N
 _b-\bar NN_b\bigr)\Bigr)+ \nonumber \\
\!\!\!\!&+\!\!\!\!&\frac{1}{2}\Bigl({}^{\cal S}R+\bigl(\nu_{ab}\nu
 ^{ab}-\nu^2-\tau_{ab}\tau^{ab}+\tau^2\bigr)\Bigr){}^\bot\varepsilon
 _{cd}K^c\bar K^d+ \nonumber \\
\!\!\!\!&+\!\!\!\!&v^a\bigl(D_aN\bigr)\Bigl(\Delta_b\bar K^b-\tau\bar 
 N\Bigr)-v^a\bigl(D_a\bar N\bigr)\Bigl(\Delta_bK^b-\tau\bar N\Bigr)+
 \nonumber \\
\!\!\!\!&+\!\!\!\!&\Bigl(2A_aA^a-\tau\chi_{ab}v^av^b\Bigr){}^\bot
 \varepsilon_{cd}K^c\bar K^d+ \nonumber \\
\!\!\!\!&+\!\!\!\!&\Bigl(\bar K^a\bigl(\delta_aK_b\bigr)-K^a\bigl(
 \delta_a\bar K_b\bigr)\Bigr)t^b\nu+ \nonumber \\
\!\!\!\!&+\!\!\!\!&A_a\Bigl(K^a\bar K^b-\bar K^aK^b\Bigr)v_b\nu+
 \nonumber \\
\!\!\!\!&+\!\!\!\!&\bigl(\nu^{ab}-\nu q^{ab}\bigr)t_c\Bigl(\bar K^c
 \delta_aK_b-K^c\delta_a\bar K_b\Bigr)+ \nonumber \\
\!\!\!\!&+\!\!\!\!&\bigl(\delta_e\bar K_a\bigr)\bigl(\delta^eK_b
 \bigr)\,{}^\bot\varepsilon^{ab}+ \nonumber \\
\!\!\!\!&+\!\!\!\!&A^b\Bigl(\bigl(\delta_b\bar K_a\bigr)v^av_cK^c-
 \bigl(\delta_bK_a\bigr)v^av_c\bar K^c\Bigr)+ \nonumber \\
\!\!\!\!&+\!\!\!\!&A^b\Bigl(\bigl(\delta_bK_a\bigr)t^at_c\bar K^c-
 \bigl(\delta_b\bar K_a\bigr)t^at_cK^c\Bigr)+ \nonumber \\
\!\!\!\!&+\!\!\!\!&A^c\tau_{ca}t_b\bigl(K^a\bar K^b-\bar K^a\bar K^b
 \bigr). \label{eq:4.2.c}
\end{eqnarray}
Adding the three integrands and forming total $\delta_a$-divergences, 
the resulting expression can be written in the form 

\begin{eqnarray}
2\kappa\bigl(I_1\!\!\!\!&+\!\!\!\!&I_2+I_3\bigr)=\Bigl(\bar K^e
 \delta_eK_a-K^e\delta_e\bar K_a\Bigr)\,{}^\bot\varepsilon^{ab}Q^c{}
 _{cb}+ \nonumber \\
\!\!\!\!&+\!\!\!\!&\bigl(\delta_e\bar K^a\bigr)\bigl(\delta^eK^b\bigr)
 \,{}^\bot\varepsilon_{ab}-\lambda\,{}^\bot\varepsilon_{ab}K^a\bar K^b
 + \nonumber \\
\!\!\!\!&+\!\!\!\!&\bigl(\delta_aA_b-\delta_bA_a\bigr)K^a\bar K^b+ 
 \nonumber \\
\!\!\!\!&+\!\!\!\!&\frac{1}{2}\Bigl({}^{\cal S}R+\tau_{ab}\tau^{ab}
 -\tau^2-\nu_{ab}\nu^{ab}+\nu^2\Bigr)\,{}^\bot\varepsilon_{cd}K^c\bar 
 K^d+ \nonumber \\
\!\!\!\!&+\!\!\!\!&\bigl(\Delta_a\bar K_b\bigr)Q^{abc}\,{}^\bot
 \varepsilon_{cd}K^d-\bigl(\Delta_aK_b\bigr)Q^{abc}\,{}^\bot
 \varepsilon_{cd}\bar K^d+ \nonumber \\
\!\!\!\!&+\!\!\!\!&\bigl(\Delta_b\bar K^b\bigr)\Bigl(v^eD_eN+v_eN^e
 \chi_{cd}v^cv^d-A_eN^e-Q^e{}_{ec}\,{}^\bot\varepsilon^{cd}K_d\Bigr)-
 \nonumber \\
\!\!\!\!&-\!\!\!\!&\bigl(\Delta_bK^b\bigr)\Bigl(v^eD_e\bar N+v_e\bar 
 N^e\chi_{cd}v^cv^d-A_e\bar N^e-Q^e{}_{ec}\,{}^\bot\varepsilon^{cd}
 \bar K_d\Bigr)+\nonumber \\
\!\!\!\!&+\!\!\!\!&\delta_a\Bigl(\bigl(\bar K^aK^b-K^a\bar K^b\bigr)
 A_b+\bigl(\nu^{ab}-\nu q^{ab}\bigr)t^c\bigl(K_c\bar K_b-\bar K_cK_b
 \bigr)\Bigr). \label{eq:4.3}
\end{eqnarray}
Since the third line is just the contraction of the curvature of the 
connection $\delta_e$ in the normal bundle and the vector fields $K^a$ 
and $\bar K^a$, moreover the fourth line is proportional to the trace 
of (\ref{eq:2.5.a}); one might attempt to rewrite (\ref{eq:4.3}) in 
a form containing the curvature $F^a{}_{bcd}$ of $\Delta_e$. We show 
that this can indeed be done. 

To get terms such as the right hand side of (\ref{eq:2.5.b}) and 
(\ref{eq:2.5.c}), let us rewrite the first as well as the fifth lines 
of (\ref{eq:4.3}) as total $\delta_e$-divergences and terms with the 
derivative of the extrinsic curvature tensor. Re-expressing the second 
line in terms of the $\Delta_e$-derivative operator we have

\begin{eqnarray}
2\kappa\bigl(I_1\!\!\!\!&+\!\!\!\!&I_2+I_3\bigr)=-\lambda\,{}^\bot
 \varepsilon_{cd}K^c\bar K^d-\bigl(\Delta^eK^a\bigr)\bigl(\Delta_e
 \bar K^b\bigr)\,{}^\bot\varepsilon_{ab}+ \nonumber \\
\!\!\!\!&+\!\!\!\!&\frac{1}{2}\Bigl({}^{\cal S}R+\nu_{ab}\nu^{ab}-
 \nu^2-\tau_{ab}\tau^{ab}+\tau^2\Bigr)\,{}^\bot\varepsilon_{cd}K^c
 \bar K^d+ \nonumber \\
\!\!\!\!&+\!\!\!\!&\frac{1}{2}\Bigl(\delta_aA_b-\delta_bA_a+\tau_a
 {}^e\nu_{eb}-\tau_b{}^e\nu_{ea}\Bigr)\bigl(K^a\bar K^b-\bar K^aK^b
 \bigr)+ \nonumber \\
\!\!\!\!&+\!\!\!\!&\Bigl(\delta_aQ^a{}_{ce}-\delta_cQ^a{}_{ae}\Bigr)
 \,{}^\bot\varepsilon^e{}_d\bigl(K^c\bar K^d-\bar K^cK^d\bigr)+
 \nonumber \\
\!\!\!\!&+\!\!\!\!&\bigl(\Delta_b\bar K^b\bigr)\Bigl(v^eD_eN+v_eN^e
 \chi_{cd}v^cv^d-A_eN^e\Bigr)-\nonumber \\
\!\!\!\!&-\!\!\!\!&\bigl(\Delta_bK^b\bigr)\Bigl(v^eD_e\bar N+v_e\bar 
 N^e\chi_{cd}v^cv^d-A_e\bar N^e\Bigr)+\nonumber \\
\!\!\!\!&+\!\!\!\!&\delta_a\Bigl(\bigl(\bar K^aK^b-K^a\bar K^b\bigr)
 \bigl(A_b+\,{}^\bot\varepsilon_{bc}Q_e{}^{ec}\bigr)+\bigl(\bar K^e
 K_b-K^e\bar K_b\bigr)\,{}^\bot\varepsilon_{ef}Q^{abf}+ \nonumber \\
\!\!\!\!&{}\!\!\!\!&+\bigl(\nu^{ab}-\nu q^{ab}\bigr)t^c\bigl(K_c\bar 
 K_b-\bar K_cK_b\bigr)\Bigr). \label{eq:4.4}
\end{eqnarray}
Comparing its second, third and fourth lines with 
(\ref{eq:2.5.a})-(\ref{eq:2.5.d}), we find that these can be rewritten 
as 

\begin{eqnarray}
&{}&\frac{1}{2}F^{ab}{}_{ab}\,{}^\bot\varepsilon_{cd}K^c\bar K^d+
 \frac{1}{2}\,{}^\bot\varepsilon_{ab}F^{ab}{}_{cd}K^c\bar K^d+F^{ab}
 {}_{bc}\,{}^\bot\varepsilon_{ad}\bigl(K^c\bar K^d-\bar K^cK^d\bigr)
 = \nonumber \\
&{}&=\frac{1}{8}K^a\bar K^bF^{cd}{}_{ef}\,{}^\bot\varepsilon_{gh}
 \delta^{efgh}_{abcd}, \label{eq:4.5}
\end{eqnarray}
by means of which we arrive at our final expression for the formal 
Poisson bracket of two constraint functions: 

\begin{eqnarray}
\Bigl\{C\bigl[N,N^a\bigr]\!\!\!\!\!\!&,\!\!\!\!\!\!&C\bigl[\bar N,
 \bar N^a\bigr]\Bigr\}=\nonumber \\
\!\!\!\!&=\!\!\!\!&C\Bigl[\bar N^eD_eN-N^eD_e\bar N\, ,\,ND^a\bar N-
 \bar ND^aN-\bigl[N,\bar N\bigr]^a\Bigr]+ \nonumber \\
\!\!\!\!&+\!\!\!\!&\frac{1}{\kappa}\oint_{\cal S}\Bigl\{\lambda\,{}
 ^\bot\varepsilon_{ab}K^a\bar K^a+\bigl(\Delta^eK^a\bigr)\bigl(
 \Delta_e\bar K^b\bigr)\,{}^\bot\varepsilon_{ab}- \nonumber \\
&{}& -\frac{1}{8}K^a\bar K^bF^{cd}{}_{ef}\,{}^\bot\varepsilon_{gh}
 \delta^{efgh}_{abcd}+ \nonumber \\
&{}& +\bigl(\Delta_bK^b\bigl)\Bigl(v^eD_e\bar N+v_e\bar N^e\chi_{cd}
 v^cv^d-\bar N^eA_e\Bigr)- \nonumber \\
&{}& -\bigl(\Delta_b\bar K^b\bigl)\Bigl(v^eD_eN+v_eN^e\chi_{cd}v^c
 v^d-N^eA_e\Bigr)\Bigr\} {\rm d}{\cal S}. \label{eq:4.6}
\end{eqnarray}
Since the curvature $F^a{}_{bcd}$ is the pull back to ${\cal S}$ of 
the spacetime curvature 2-form ${}^MR^a{}_{bcd}$, in the physically 
important special case $n=3$ the curvature term reduces to $\frac{1}
{4}K^a\bar K^b\varepsilon_{abcd}$ ${}^MR^{cd}{}_{ef}\varepsilon^{ef}$, 
and hence gives a Penrose-type charge integral of the spacetime 
curvature \cite{PR}. In a GHP spin frame $(o^A,\iota^A)$ adapted to 
${\cal S}$ this takes the form $K^a\bar K^b\varepsilon_{A'B'}(\Psi
_{ABCD}o^C\iota^D-\Phi_{ABC'D'}\bar o^{C'}\bar\iota^{D'}+\Lambda(o_A
\iota_B+\iota_Ao_B))+ c.c.$. Thus, in particular, it is only the 
$\Psi_1$, $\Psi_2$ and the $\Psi_3$, but not the $\Psi_0$ and $\Psi
_4$ Weyl spinor components that are involved in the Poisson boundary 
term.

\subsubsection{Boundary conditions from the gauge invariance of the 
Poisson boundary term}
\label{sub-4.1.4}

Clearly, the first three terms in the boundary integral of 
(\ref{eq:4.6}) are manifestly $2+(n-1)$-covariant; they depend only 
on the geometry of ${\cal S}$ and the value of the vector fields on 
${\cal S}$ (but independent of the way in which they are extended off 
the boundary), and they are invariant with respect to the change of 
the actual normals $(t^a,v^a)$ of ${\cal S}$. On the other hand, the 
last two lines contain `bad' terms. Thus, we can ensure the gauge 
invariance of the Poisson boundary term (in the sense discussed in 
subsection \ref{sub-4.1.2}) if \emph{we require the vanishing of the 
$\Delta_e$-divergence of the vector fields} $K^a$ and $\bar K^a$. 
Obviously, $\Delta_aK^a=0$ is a $2+(n-1)$ covariant condition, and 
from $\Delta_aK^a=\delta_aK^a+Q^a{}_{ab}K^b=N\tau-v_aN^a\nu+\delta_a
(\Pi^a_bN^b)$ it is clear that it has infinitely many solutions on 
${\cal S}$: the condition $\Delta_aK^a=0$ specifies e.g. only the 
lapse $N$ in terms of the still completely freely specifiable shift 
$N^a$. 

To see the meaning of this condition, let us rewrite this into the 
form $q^{ab}\Pi^c_a\Pi^d_b$ $\nabla_{(c}K_{d)}=0$. This is one of the
$\frac{1}{2}n(n-1)$ projected parts of the $\frac{1}{2}(n+1)(n+2)$ 
spacetime Killing equations. Thus, $\Delta_aK^a=0$ is a weakening of 
the familiar spacetime Killing equations. Clearly, if the lapse part 
of $K^a$ is vanishing and the shift part $N^a$ is tangent to ${\cal 
S}$ on ${\cal S}$, then $\Delta_aK^a=0$ reduces to the condition 
$\delta_aN^a=0$ discussed in subsection \ref{sub-2.3.2}. 

To clarify its compatibility with the evolution equations, let us 
rewrite the canonical equation of motion for the metric $h_{ab}$ in 
the spacetime. By (\ref{eq:3.13.c}), its right-hand side is just the 
projection to $\Sigma$ of the spacetime Killing operator: 

\begin{equation}
\dot h_{ab}=2N\chi_{ab}+L_{\bf N}h_{ab}=2P^c_aP^d_b\nabla_{(c}K_{d)}.
\nonumber
\end{equation}
Hence, the contraction of its restriction to ${\cal S}$ with the metric 
$q^{ab}$ gives $q^{ab}\dot h_{ab}=2\Delta_bK^b$. However, the left-hand 
side is proportional to the time derivative of the induced volume 
element on ${\cal S}$: $\dot\varepsilon_{e_1\dots e_{n-1}}=\frac{1}{2}
q^{ab}\dot q_{ab}\varepsilon_{e_1\dots e_{n-1}}=\frac{1}{2}q^{ab}\dot 
h_{ab}\varepsilon_{e_1\dots e_{n-1}}$. Therefore, $\Delta_aK^a=0$ is 
precisely the condition that \emph{the induced volume $(n-1)$-form on 
${\cal S}$ is constant during the evolution}. Thus the boundary 
condition $\delta\varepsilon_{e_1\dots e_{n-1}}=0$ for the 
configuration variables, found in the special cases and discussed in 
subsections \ref{sub-2.3.1} and \ref{sub-2.3.2}, appears naturally in 
the general case, too.

\subsubsection{On Einstein--scalar systems}
\label{sub-4.1.5}

The quasi-local phase space of the coupled Einstein-scalar system is 
the cotangent bundle $T^*{\cal Q}(\Sigma)$ of the configuration space 
${\cal Q}(\Sigma)$, the latter being the set of the pairs $(h_{ab},
\Phi)$, and endowed with the natural symplectic structure. The 
constraint of the coupled system is $C[N,N^a]:=C_E[N,N^a]+H_S[N,N^a]
=0$, where now (\ref{eq:1.1}) is denoted by $C_E[N,N^a]$ and $H_S[N,
N^a]$ is given by (\ref{eq:3.4}), and the Hamiltonian is the sum of 
the Hamiltonians of the gravitational and the scalar sectors: $H[N,N^a]
=H_E[N,N^a]+H_S[N,N^a]$. However, note that $H_S[N,N^a]$ depends on 
the metric $h_{ab}$, which is now a configuration variable. Thus, 
assuming that both $K^a:=Nt^a+N^a$ and $\bar K^a:=\bar Nt^a+\bar N^a$ 
are $\Delta_e$--divergence-free, the \emph{formal} Poisson bracket of 
the Hamiltonians $H[N,N^a]$ and $H[\bar N,\bar N^a]$ is 

\begin{eqnarray}
\Bigl\{H\bigl[N,N^a\bigr]\!\!\!\!\!\!&,\!\!\!\!\!\!&H\bigl[\bar N,
 \bar N^a\bigr]\Bigr\}=\label{eq:4.8} \\
\!\!\!\!&=\!\!\!\!&C\Bigl[\bar N^eD_eN-N^eD_e\bar N\, ,\,ND^a\bar N-
 \bar ND^aN-\bigl[N,\bar N\bigr]^a\Bigr]+ \nonumber \\
\!\!\!\!&+\!\!\!\!&\frac{1}{\kappa}\oint_{\cal S}\Bigl\{\bigl(\Delta
 ^eK^a\bigr)\bigl(\Delta_e\bar K^b\bigr)\,{}^\bot\varepsilon_{ab}+
 \lambda\,{}^\bot\varepsilon_{ab}K^a\bar K^a- \nonumber \\
&{}& -\frac{1}{8}K^a\bar K^bF^{cd}{}_{ef}\,{}^\bot\varepsilon_{gh}
 \delta^{efgh}_{abcd}+\frac{1}{2}\bigl(K^a\bar K^b-\bar K^aK^b\bigr)
 \,{}^\bot\varepsilon_{ac}\kappa T^c{}_b\Bigr\} {\rm d}{\cal S}. 
 \nonumber
\end{eqnarray}
Since $F^a{}_{bcd}={}^MR^a{}_{bef}\Pi^e_c\Pi^f_d$, the last three 
terms of the boundary integral can be written as 

\begin{eqnarray}
&{}&-\frac{1}{8}K^a\bar K^b{}^MC^{cd}{}_{ef}\,{}^\bot\varepsilon_{gh}
 \delta^{efgh}_{abcd}+\nonumber \\
&{}&+\frac{1}{2}\bigl(K^a\bar K^b-\bar K^aK^b\bigr)\,{}^\bot
 \varepsilon_{ac}\Bigl(\,{}^MG^c{}_b+\kappa T^c{}_b+\delta^c_b\lambda
 \Bigr)+\nonumber \\
&{}&+\frac{1}{2}\bigl(K^a\bar K^b-\bar K^aK^b\bigr)\,{}^\bot
 \varepsilon_{ac}\Bigl(\frac{n-3}{n-1}\,{}^MR^c{}_b+\frac{1}{n(n-1)}
 \,{}^MR\delta^c_b\Bigr), \nonumber
\end{eqnarray}
where ${}^MC_{abcd}$, ${}^MR_{ab}$ and ${}^MR$ are the spacetime Weyl 
and Ricci tensors and the curvature scalar, respectively, and the 
second line is proportional to the expression whose vanishing is just 
the Einstein equation. Thus, in particular for $n=3$ and `on shell', 
the Poisson bracket of two Hamiltonians is the boundary integral of 
$(\Delta^eK^a)(\Delta_e\bar K^b)\,{}^\bot\varepsilon_{ab}-\frac{1}{2}
K^a\bar K^b\,{}^MC_{abcd}\,{}^\bot\varepsilon^{cd}+\frac{1}{6}{}^MR
K^a\bar K^b\,{}^\bot\varepsilon_{ab}$. Hence, the trace-free part of 
the spacetime Ricci tensor does not appear even in the presence of a 
scalar field. 


\subsection{Quasi-local quantities from Poisson boundary terms?}
\label{sub-4.2}

In subsection \ref{sub-4.1.2} we raised the idea that the Poisson 
boundary term should be the sum of the Hamiltonian boundary term 
(parameterized by the new lapse and shift) and terms analogous to the 
flux of energy-momentum/angular momentum of matter fields, and both 
must be gauge invariant. In the present subsection, we decompose the 
(gauge-invariant) Poisson boundary term in such a way that the 
boundary term of the `improved' basic Hamiltonian (\ref{eq:2.8}) 
emerges naturally, even in a (slightly modified) gauge-invariant form. 
However, as we will see, in its gauge-invariant form it does not seem 
to yield a representation of the `composition rule' of how the new 
lapse and shift are built from the old ones. 

We start with (\ref{eq:4.3}), and let us observe first that in its 
first line the derivative operator $\delta_e$ can be replaced by 
$\Delta_e$; moreover, the first term in the second line can be written 
as $\frac{1}{2}(\delta_e\delta^eK^a)\,{}^\bot\varepsilon_{ab}\bar K^b
-\frac{1}{2}(\delta_e\delta^e\bar K^a)\,{}^\bot\varepsilon_{ab}K^b$ up 
to a total $\delta_e$-divergence. Again, by forming total 
$\delta_a$-divergences, the third line is written as $(\bar K^e\Delta
_eK^a-K^e\Delta_e\bar K^b)A_b$ plus extrinsic curvature terms. Thus, 
for vector fields satisfying $\Delta_aK^a=\Delta_a\bar K^a=0$, we 
obtain 

\begin{eqnarray}
\Bigl\{H\bigl[N,N^a\bigr]\!\!\!\!\!\!&,\!\!\!\!\!\!&H\bigl[\bar N,
 \bar N^a\bigr]\Bigr\}=\label{eq:4.9} \\
\!\!\!\!&=\!\!\!\!&C\Bigl[\bar N^eD_eN-N^eD_e\bar N\, ,\,ND^a\bar N-
 \bar ND^aN-\bigl[N,\bar N\bigr]^a\Bigr]- \nonumber \\
\!\!\!\!&-\!\!\!\!&\frac{1}{\kappa}\oint_{\cal S}\Bigl\{\bigl(\bar K^e
 \Delta_eK_a-K^e\Delta_e\bar K_a\bigr)\Bigl(\,{}^\bot\varepsilon
 ^{ab}Q^c{}_{cb}+A^a\Bigr)+ \nonumber \\
&{}& +\frac{1}{2}\Bigl(\delta_e\delta^eK^a-2\bigl(\delta_cK_d\bigr)
 Q^{cda}-2K^eQ_{cde}Q^{cda}- \nonumber \\
&{}& \quad -K^c\bigl(Q_{cfd}-q_{cf}Q^e{}_{ed}\bigr)A^f\,{}^\bot
 \varepsilon^{da}+ \nonumber \\
&{}&\quad +\frac{1}{2}K^a\bigl(\,{}^MR-2\lambda+\bigl[Q_{cde}-q_{cd}
 Q^f{}_{fe}\bigr]Q^{cde}\bigr)\Bigr)\,{}^\bot\varepsilon_{ab}\bar 
 K^b-\nonumber \\
&{}& -\frac{1}{2}\Bigl(\delta_e\delta^e\bar K^a-2\bigl(\delta_c\bar 
 K_d\bigr)Q^{cda}-2\bar K^eQ_{cde}Q^{cda}- \nonumber \\
&{}& \quad -\bar K^c\bigl(Q_{cfd}-q_{cf}Q^e{}_{ed}\bigr)A^f\,{}^\bot
 \varepsilon^{da}+ \nonumber \\
&{}&\quad +\frac{1}{2}\bar K^a\bigl(\,{}^MR-2\lambda+\bigl[Q_{cde}-
 q_{cd}Q^f{}_{fe}\bigr]Q^{cde}\bigr)\Bigr)\,{}^\bot\varepsilon_{ab}
 K^b\Bigr\}{\rm d}{\cal S}. \nonumber 
\end{eqnarray}
Thus the first line in the boundary integral is just the boundary term 
of $H_1[K^a]$ given by (\ref{eq:2.8}), in which $K^a$ is replaced by 
the `commutator' $\bar K^e\Delta_eK^a-K^e\Delta_e\bar K^a$. Since, 
assuming $\Delta_aK^a=\Delta_a\bar K^a=0$, the integral of 
(\ref{eq:4.3}) is invariant with respect to the change of the basis 
$(t^a,v^a)$ in the normal bundle of ${\cal S}$ (`$SO(1,1)$ boost-gauge 
invariance'), the decomposition of the integrand of (\ref{eq:4.9}) to 
the Hamiltonian boundary term (the first line) and to the rest can be 
made in a boost-gauge-invariant way, too. In fact, by the Hodge 
decomposition (see e.g. \cite{Wa}) the connection 1-form $A_e$ is the 
sum of an exact, a co-exact and a harmonic 1-form on ${\cal S}$: $A_e
=\delta_e\alpha+\alpha_e+\omega_e$, respectively, and this decomposition 
is unique. Here the function $\alpha:{\cal S}\rightarrow\mathbb{R}$ is 
unique up to an additive constant, $\alpha_e$ is $\delta_e$-divergence 
free, while $\omega_e$ is both $\delta_e$-divergence free and 
satisfies $\delta_{[a}\omega_{b]}=0$. (The first represents the pure 
gauge part of $A_e$, the co-exact part yields curvature, and the 
harmonic part only holonomy, but no curvature.) Thus by $\delta_{[a}
A_{b]}=\delta_{[a}\alpha_{b]}$ it is only the co-exact part of $A_e$ 
that appears in (\ref{eq:4.3}), and hence we can substitute $A_e$ by 
$\alpha_e$ in (\ref{eq:4.9}), too, yielding a manifestly boost gauge 
invariant form of the `improved' basic Hamiltonian. Another 
interpretation of the above Hodge decomposition is that it provides a 
`natural' gauge fixing, using only the intrinsic geometry of ${\cal 
S}$. 

Unfortunately, however, this Hamiltonian boundary term does not seem 
to represent the `composition rule' of the lapses and shifts in a 
correct way. Indeed, the lapse-shift parts of $\bar K^e\Delta_eK^a-
K^e\Delta_e\bar K^a$ are \emph{not} the new lapse $\bar N^eD_eN-N^e
D_e\bar N$ and the new shift $ND^a\bar N-\bar ND^aN-[N,\bar N]^a$ 
that appear in the constraint function.


\section{Conclusions}
\label{sec-5}

We learnt from the quasi-local canonical formulation of the scalar 
field that the Poisson boundary terms represent energy-momentum and 
angular momentum fluxes out from and into the localized system, i.e. 
they are well defined physical quantities. Here we raise the idea that 
the same may be expected in general relativity, too, and hence the 
Poisson boundary term in GR must be gauge invariant in every sense. 
We showed that this requirement yields the condition for the lapse 
and shift that the spacetime vector field that they determine must be 
divergence free with respect to a Sen-type connection on the boundary. 
This condition is a part of the spacetime Killing equations. This 
yields that the evolution equations preserve the volume form induced 
on the boundary. Therefore, keeping the induced volume form fixed as 
a condition seems to be the part of the (yet unknown) `ultimate' 
boundary conditions for the canonical variables. This implies that the 
quasi-local constraint algebra that we found earlier (and discussed in 
subsection \ref{sub-2.3.1}) is probably the `correct' one, completing 
the point (iii) of subsection \ref{sub-2.1.1}. 

We also found arguments both in favour of and against the Hamiltonian 
boundary term in $H_1[K^a]$. It appears naturally as a part of the 
Poisson boundary term and its $SO(1,1)$ boost-gauge dependence can be 
cured, but, without additional restrictions on the lapse and the shift 
it does not yield functionally differentiable Hamiltonian, and it does 
not represent the composition law for the lapses and shifts in a 
correct way.


\bigskip

The author is grateful to Robert Beig, J\"org Frauendiener, Helmut 
Friedrich and Niall \'O Murchadha for the useful remarks and 
discussions on the canonical formulation of general relativity and the 
role of boundary conditions, to Edward Anderson and the referees for 
their bibliographic remarks, as well as the Erwin Schr\"odinger 
Institute, Vienna, for hospitality. This work was partially supported 
by the Hungarian Scientific Research Fund (OTKA) grant K67790.

\end{document}